# Radiative symmetry breaking with a scale invariant seesaw mechanism

Aqeel Ahmed,* Juan P. Garcés,† and Manfred Lindner‡
*Max-Planck-Institut für Kernphysik, Saupfercheckweg 1, 69117 Heidelberg, Germany*

We study a scale invariant inverse seesaw model that radiatively generates the electroweak scale, the Standard Model (SM) neutrino masses, and stabilizes the electroweak vacuum. Previous studies have noted that the SM Higgs potential and the electroweak scale can be radiatively generated via the minimal seesaw mechanism. This scenario, called the "neutrino option," was UV completed by a scale invariant framework. However, these models predict singlet neutrino and scalar masses at $10^7$–$10^9$ GeV, beyond any experimental reach and leave the electroweak vacuum metastable/unstable at high energies. In this work, we propose modifications to this framework that lower the singlet neutrino masses to experimentally accessible scales through an inverse seesaw mechanism, while fully stabilizing the electroweak vacuum with an additional singlet scalar. The possibility of generating the observed baryon asymmetry of the Universe via leptogenesis is also explored.

## I. INTRODUCTION

The SM is an exceptionally successful theory in particle physics, rigorously validated through precise experimental tests. However, it falls short in explaining phenomena such as dark matter, dark energy, neutrino masses, and the baryon asymmetry of the Universe. Furthermore, the peculiar numerical values and structures required to explain the SM fermion masses and mixings (*flavor puzzle*) and the light Higgs mass (*hierarchy problem*) motivate the need for extensions that aim to address these deficiencies and introduce a sense of naturalness while maintaining minimal complexity.

At the classical level, the SM is an almost conformal/scale invariant theory; the only source of scale symmetry breaking is the Higgs boson mass parameter.[1] Therefore, it seems natural to consider models where the SM Higgs mass parameter is absent at high-energy scales and is radiatively generated at low scales. In this context, it has been recently pointed out that the minimal seesaw mechanism [1–3], which manifestly generates masses for the light SM neutrinos, can also radiatively generate the Higgs potential through loops containing heavy singlet right-handed neutrinos (RHNs) [4], dubbed as the "neutrino option" (see also [5,6]). The viability of such a model with a vanishing tree-level Higgs mass parameter prompted an exploration of a conformal embedding, where the RHN mass scale is generated via spontaneous symmetry breaking of the classical conformal symmetry. In this context, a minimal classically conformal extension of the neutrino option with two real scalar singlets was proposed in [7], achieving consistent outcomes while maintaining perturbativity and scale invariance up to the Planck scale.[2]

Despite these improvements, some shortcomings persist. Foremost, the predicted heavy neutrino masses lie in the range of 10–500 PeV, rendering the model currently untestable. Furthermore, the Higgs self-coupling in the SM exhibits negative renormalization group (RG) running and enters a region of *metastability* at high energies $\mathcal{O}(10^{10})$ GeV [9–11]. In both realizations of the neutrino option mentioned above, the corrections to the Higgs self-coupling are significantly suppressed, and its RG running is similar to that of the SM. Therefore, the Higgs potential remains unstable/metastable at high energies, as in the SM.

In this study, we address these shortcomings firstly, by lowering the RHN masses $M_N$ to lie in the range $(1-10^6)$ GeV, which includes the experimentally testable range $\mathcal{O}(1)$ GeV $\lesssim M_N \lesssim \mathcal{O}(10^2)$ GeV [12] and secondly, by seeking sizable shifts to the RG running of the Higgs self-coupling via the Higgs portal, such that the Higgs scalar potential is rendered fully stable. All of this is achieved in a minimal extension of the SM while preserving UV scale invariance and perturbativity, and the correct light neutrino masses. More specifically, we modify the model presented in [7] so that the type-I seesaw framework is replaced by an inverse seesaw (IS) framework [13] (see also [14]). The nonvanishing vacuum expectation values (VEVs) of the beyond the Standard Model (BSM) scalars are dynamically generated via spontaneously broken conformal symmetry (SBCS) *à la* Coleman-Weinberg [15], and serve as the required scales within the minimal inverse seesaw framework. Furthermore, a global U(1) symmetry (lepton number) is implemented to achieve the inverse seesaw mass matrix texture, yet it is subsequently spontaneously broken by a Yukawa term that induces a Majorana mass upon the breaking of scale invariance. With this, introducing a small coupling associated with the Majorana mass term preserves naturalness in the t'Hooft sense [16].

*Contact author: aqeel.ahmed@mpi-hd.mpg.de
†Contact author: juan.garces@mpi-hd.mpg.de
‡Contact author: lindner@mpi-hd.mpg.de

[1]At the quantum level, scale symmetry is anomalously broken by the renormalization group equation (RGE) running of the SM couplings.

[2]The necessity of at least two real scalars for a UV-complete conformal realization is explained in [8].

The model parameter space is defined at the SBCS scale, and the PyR@TE3 package [17] is employed to compute one-loop RGEs, which are then used to explore the low- and high-energy behaviors of a given set of boundary conditions. The parameter space is constrained by requiring the correct Higgs mass and Higgs self-coupling at the top mass scale in the modified minimal subtraction scheme $\overline{\mathrm{MS}}$, agreement with observed light neutrino masses, full UV perturbativity, and classical scale invariance up to the Planck scale.[3] Additional constraints stem from requiring a fully stable Higgs scalar potential. Furthermore, the possibility of realizing the observed baryon asymmetry of the Universe in this scenario is also explored; see also [18,19] for leptogenesis discussion in the context of the neutrino option. However, this is done in a rather simplistic fashion, and a more rigorous study, including the numerical evaluation of the Boltzmann equations together with the UV RG running of the couplings, will be the subject of a future study.

The outline of the paper is as follows: In Sec. II, we provide details of the minimal model, including the scale invariant setup, the realization of the inverse seesaw framework below the SBCS scale, and the computation of field-dependent masses along with the associated one-loop threshold corrections required for matching the model to its effective low-energy version. In Sec. III, we perform a numerical analysis and present the model observables compared to the respective experimental constraints. Finally, in Sec. IV, we introduce a third scalar singlet to construct a scale invariant model that achieves full vacuum stability and correct low-energy behavior. We summarize and conclude the paper in Sec. V.

[3]It is expected that quantum gravity becomes relevant at the Planck scale. Also, note that this scale could be lowered to the scale of grand unified theories (GUT) in the case of a Grand Unified conformal embedding. The Planck scale is chosen here because parameter space points that satisfy all the constraints, in this case, would immediately form a subset of the model's parameter space with a GUT embedding.

TABLE I. Relevant particle content and corresponding quantum numbers under the electroweak gauge group and a global lepton number $U(1)_L$. All new particles are singlets under the color gauge group of the SM. Here, we suppress the fermion generation indices for brevity.

| Particle | Spin | $SU(2)_L \otimes U(1)_Y$ | $U(1)_L$ |
|---|---|---|---|
| $L$ | 1/2 | $(\mathbf{2},-1/2)$ | +1 |
| $N_R$ | 1/2 | $(\mathbf{1},0)$ | +1 |
| $\xi_R$ | 1/2 | $(\mathbf{1},0)$ | −1 |
| $H$ | 0 | $(\mathbf{2},-1/2)$ | 0 |
| $S$ | 0 | $(\mathbf{1},0)$ | 0 |
| $\hat{\phi}$ | 0 | $(\mathbf{1},0)$ | +2 |

## II. THE MODEL

We consider a scale-invariant version of the SM that includes one real and one complex scalar singlet, denoted $S$ and $\hat{\phi}$, respectively, along with two pairs of Weyl fermion singlets $N_R^i$ and $\xi_R^i$ ($i = 1, 2$).[4] Furthermore, a global $U(1)_L$ lepton number is assigned to these particles, as specified in Table I. The model Lagrangian is given by

$$\begin{aligned}\mathcal{L} \supset & \frac{1}{2}\partial_\mu S \partial^\mu S + (\partial_\mu \hat{\phi})^* \partial^\mu \hat{\phi} - V(H, S, \phi) \\ & + i\bar{N}_R \partial\!\!\!/ N_R + i\bar{\xi}_R \partial\!\!\!/ \xi_R - y_\nu \bar{L}\,\tilde{H}\,N_R \\ & - y_N S N_R^T C^{-1} \xi_R - \frac{1}{2} y_M \hat{\phi} \xi_R^T C^{-1} \xi_R + \mathrm{H.c.}, \end{aligned} \tag{1}$$

where $L$ denotes the SM lepton doublets, $H$ is the SM Higgs doublet (with its conjugate field $\tilde{H} \equiv i\sigma_2 H^*$), and $C \equiv i\gamma^2\gamma^0$ represents the charge conjugation matrix. The above $y$'s are dimensionless Yukawa couplings.

The most general gauge-invariant scalar potential $V(H, S, \phi)$, respecting scale symmetry and the global $U(1)_L$ specified above, is given by

$$\begin{aligned}V(H, S, \phi) = & \lambda |H|^4 + \lambda_S S^4 + \lambda_{\hat{\phi}} |\hat{\phi}|^4 + \lambda_{HS} |H|^2 S^2 \\ & + \lambda_{H\hat{\phi}} |H|^2 |\hat{\phi}|^2 + \lambda_{S\hat{\phi}} S^2 |\hat{\phi}|^2, \end{aligned} \tag{2}$$

where $\lambda$'s are dimensionless quartic couplings. In the following, we parametrize the Higgs doublet and the complex scalar as

$$H = \frac{1}{\sqrt{2}} \begin{pmatrix} \sqrt{2}\pi^+ \\ h + i\pi^0 \end{pmatrix}, \qquad \hat{\phi} = \frac{\phi}{\sqrt{2}} e^{i\sigma}, \tag{3}$$

where $\pi$'s and $\sigma$ are the would-be Goldstone bosons arising after spontaneous symmetry breaking. With the above parametrization, the scalar potential takes the form

[4]Note that the minimal conformal extension of the neutrino option [7] employed two real scalars and a RHN $N_R$ for the type-I seesaw framework.

$$V(H,S,\phi) = \lambda|H|^4 + \lambda_S S^4 + \lambda_\phi \phi^4 + \lambda_{HS}|H|^2 S^2 + \lambda_{H\phi}|H|^2\phi^2 + \lambda_{S\phi}S^2\phi^2, \tag{4}$$

where $\lambda_\phi = \lambda_{\hat{\phi}}/4, \lambda_{H\phi} = \lambda_{H\hat{\phi}}/2$, and $\lambda_{S\phi} = \lambda_{S\hat{\phi}}/2$.

Next, we focus on the three massless scalar degrees of freedom of the above potential that could potentially develop a VEV. We parametrize them collectively as

$$\vec{\varphi} \equiv (h, S, \phi). \tag{5}$$

At tree level, the above potential has a trivial vacuum structure; however, taking into account the radiative corrections, a nontrivial vacuum is generated. To obtain the nontrivial vacuum through radiative corrections, in the following, we employ the Gildener-Weinberg (GW) [20] approach. The GW approach relies on the fact that at high scale, denoted as $\Lambda_{\rm GW}$, a flat direction is achieved through the running of the quartic couplings. Along this direction, radiative corrections dynamically generate a nontrivial vacuum *à la* Coleman-Weinberg [15]; see also [21]. In particular, for our potential above, we note that for a flat direction of the form

$$\langle\vec{\varphi}\rangle \equiv (\langle h\rangle, \langle S\rangle, \langle\phi\rangle) = \varphi(0, \cos\alpha, \sin\alpha), \tag{6}$$

the GW relations that must be satisfied at the conformal symmetry breaking scale $\Lambda_{\rm GW}$ are[5]

$$\lambda_{S\phi} = -\frac{2\lambda_S}{\tan^2\alpha}\bigg|_{\Lambda_{\rm GW}}, \qquad \lambda_\phi = \frac{\lambda_S}{\tan^4\alpha}\bigg|_{\Lambda_{\rm GW}}, \tag{7}$$

where the angle $\alpha$ quantifies the hierarchy between the two nonvanishing VEVs generated at $\Lambda_{\rm GW}$, i.e.,

$$\tan\alpha \equiv \frac{v_\phi}{v_S} = \frac{\langle\phi\rangle}{\langle S\rangle}\bigg|_{\Lambda_{\rm GW}}. \tag{8}$$

Note that an important assumption that follows from choosing such a flat direction is that the Higgs field has no VEV at the GW scale, i.e., $\langle h\rangle|_{\Lambda_{\rm GW}} = 0$.

Below the GW scale, both $S$ and $\phi$ become massive. Their corresponding (field-dependent) mass terms, obtained from Eq. (4), can be concisely written as

$$\mathcal{L} \supset -\frac{1}{2}(\phi \quad S)\begin{pmatrix} m_\phi^2(h) & 4\lambda_{S\phi}v_S v_\phi \\ 4\lambda_{S\phi}v_S v_\phi & m_S^2(h)\end{pmatrix}\begin{pmatrix}\phi \\ S\end{pmatrix}, \tag{9}$$

where the diagonal entries of the mass matrix are defined as

$$m_\phi^2(h) \equiv 12\lambda_\phi v_\phi^2 + \lambda_{H\phi}h^2 + 2\lambda_{S\phi}v_S^2, \\ m_S^2(h) \equiv 12\lambda_S v_S^2 + \lambda_{HS}h^2 + 2\lambda_{S\phi}v_\phi^2. \tag{10}$$

We diagonalize the above symmetric matrix with an orthogonal transformation $U$ parametrized by a single mixing angle $\theta$ to obtain the mass eigenstates $\phi_\pm$. The corresponding tree-level field-dependent masses are

$$m_\pm^2(h) = \frac{1}{2}m_\phi^2(h) + \frac{1}{2}m_S^2(h) \pm \frac{1}{2}\sqrt{(m_\phi^2(h) - m_S^2(h))^2 + 64\lambda_{S\phi}^2 v_S^2 v_\phi^2}. \tag{11}$$

At the GW scale $\Lambda_{\rm GW}$, the GW relations (7) hold and can be used to simplify the above expression as

$$m_+^2(h) = 8\frac{\lambda_S}{\sin^2\alpha}v_S^2 + \frac{1}{2}\Big(\lambda_{H\phi} + \lambda_{HS} - (\lambda_{HS} - \lambda_{H\phi})\cos 2\alpha\Big)h^2 + \frac{\cos^2\alpha\sin^4\alpha}{8\lambda_S v_S^2}(\lambda_{H\phi} - \lambda_{HS})^2 h^4 + \cdots, \tag{12}$$

$$m_-^2(h) = \frac{1}{2}\Big(\lambda_{H\phi} + \lambda_{HS} + (\lambda_{HS} - \lambda_{H\phi})\cos 2\alpha\Big)h^2 - \frac{\cos^2\alpha\sin^4\alpha}{8\lambda_S v_S^2}(\lambda_{H\phi} - \lambda_{HS})^2 h^4 + \cdots, \tag{13}$$

where ellipses denote terms of order $\mathcal{O}(h^6)$ or higher, which are neglected. As expected, the tree-level mass of the scalon $\phi_-$ vanishes for $\langle h\rangle = 0$, and the one-loop correction of Eq. (B5) becomes relevant and must be added. Taking this into account, the field-dependent masses for the mass eigenstates, which we name $S$ and $\phi$, at the GW scale are

$$m_S^2(h) = 8B(v_S^2 + v_\phi^2) + m_-^2(h), \\ m_\phi^2(h) = m_+^2(h), \tag{14}$$

where $B$ is defined in (B3). Before discussing the threshold corrections to the Higgs potential in our conformal neutrino option, we briefly discuss the inverse seesaw mechanism and the generation of light SM neutrino masses.

### A. Inverse seesaw

In this subsection, we discuss the generation of SM neutrino masses through the IS mechanism [13,23,24]. Below the conformal symmetry-breaking scale, the terms responsible for generating light neutrino masses from (1) are

[5]For more details see Appendix B, specifically Eqs. (B11), (B12), and (B13) and the reference [22].

$$\mathcal{L} \supset (\nu_L^c)^T C^{-1} M_D N_R + N_R^T C^{-1} M_N \xi_R + \frac{1}{2} \xi_R^T C^{-1} \mu \xi_R + \text{H.c.}, \tag{15}$$

where the masses are defined as

$$M_D = \frac{y_\nu v}{\sqrt{2}}, \qquad M_N = y_N v_S, \qquad \mu = \frac{y_M v_\phi}{\sqrt{2}}, \tag{16}$$

with $v \equiv \langle h \rangle$ denoting the Higgs boson VEV. Note that the term proportional to $\mu$ in Eq. (15) spontaneously breaks the lepton number when $\phi$ acquires a nonvanishing VEV. Furthermore, the smallness of $\mu$, required to realize the inverse seesaw framework, can be achieved by assuming a small $y_M$ coupling and/or a small $v_\phi$. Since the term proportional to $y_M$ is the only one that violates the lepton number, the smallness of this coupling can still be rendered natural in the 't Hooft sense. On the other hand, the smallness of $v_\phi$ depends on the smallness of the angle $\alpha$, which parametrizes the flat direction [see Eq. (8)].

In the basis $(\nu_L^c, N_R, \xi_R)^T$, the neutrino mass matrix reads as

$$M_\nu = \begin{pmatrix} 0 & M_D & 0 \\ M_D^T & 0 & M_N \\ 0 & M_N^T & \mu \end{pmatrix}. \tag{17}$$

It is straightforward to show that, under the assumption $\mu \ll M_D \ll M_N$,[6] the leading-order contribution to the light neutrino masses from $M_\nu$ reads as

$$m_\nu = M_D (M_N^{-1})^T \mu M_N^{-1} M_D^T. \tag{18}$$

Note that in this context, the light SM neutrino masses can be naturally obtained for $\mu \ll M_N$ and $1 \lesssim M_N/\text{GeV} \lesssim 10^2$. In this case, and in striking contrast to the type-I seesaw case, the heavy singlet neutrinos can be probed at various current experiments (see below).

### B. Threshold corrections to the Higgs potential

In this subsection, we present the threshold corrections to the Higgs potential from the scalar and fermionic sectors.

#### 1. *Scalar sector*

Below the scale $\Lambda_{\text{GW}}$, where the scalar fields $S$ and $\phi$ acquire nonvanishing VEVs, tree-level corrections to the Higgs mass parameter arise from the following diagrams:

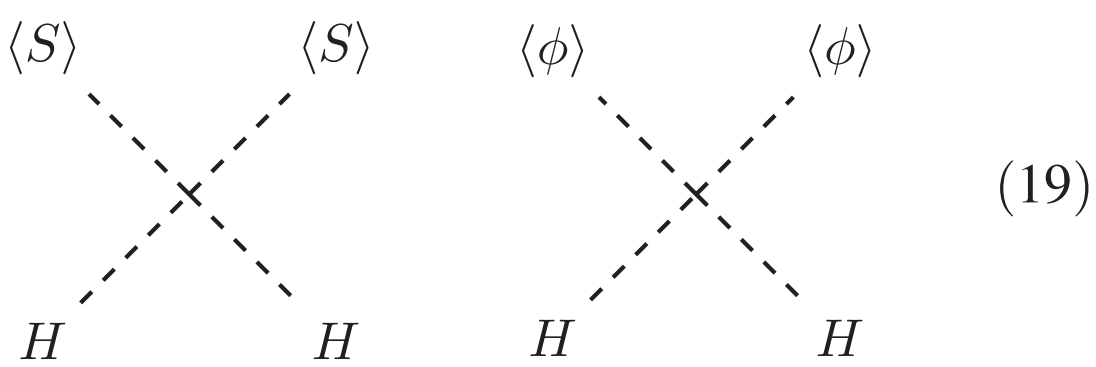

(19)

The resulting shift in the Higgs mass is given by

$$\Delta m^2_{\text{tree}} = -\lambda_{HS} v_S^2 - \lambda_{H\phi} v_\phi^2, \tag{20}$$

where we follow the convention

$$\mathcal{L} \supset m_H^2 H^\dagger H \supset \frac{1}{2} m_H^2 h^2. \tag{21}$$

These negative tree-level contributions to $m_H^2$ must generally be suppressed to achieve the correct order of magnitude for the Higgs mass at low energies. Similarly, the tree-level threshold corrections to the Higgs self-coupling stem from the following diagrams:

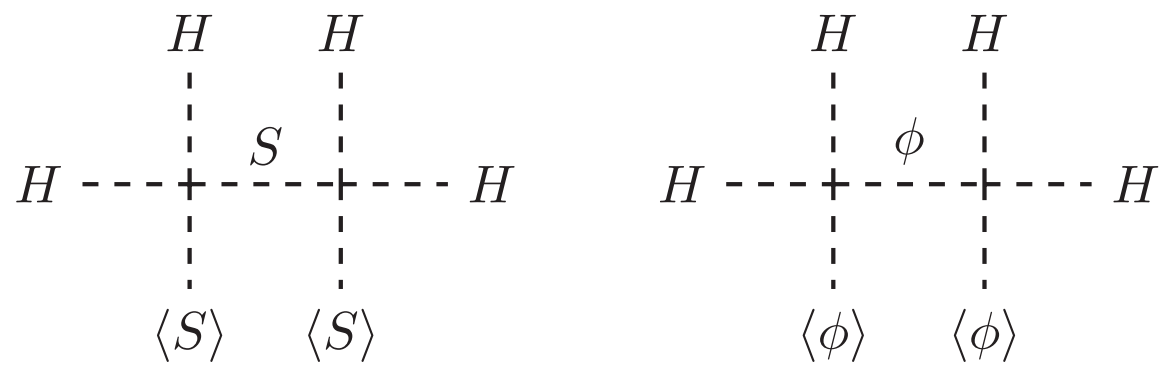

At $\Lambda_{\text{GW}}$, these tree-level shifts read as

$$\Delta\lambda_{\text{tree}} = \frac{2\lambda_{HS}^2 v_S^2}{m_S^2} + \frac{2\lambda_{H\phi}^2 v_\phi^2}{m_\phi^2}. \tag{22}$$

In order to have relatively small contributions to the Higgs mass parameter (20), very small portal couplings are needed.[7] However, for our choice of $v_\phi \ll v_S$, the suppression on $\lambda_{H\phi}$ is relaxed compared to $\lambda_{HS}$. Thus, the prospects for sizable shifts in the Higgs quartic $\lambda$ are more promising than in Ref. [7], where $\Delta m^2_{\text{tree}} = -\lambda_{HS} v_S^2$. At $\Lambda_{\text{GW}}$, the GW conditions (7) can be employed to relate the Higgs quartic contributions as

$$\Delta\lambda_{\text{tree},S} = \Delta\lambda_{\text{tree},\phi}|_{\Lambda_{\text{GW}}}, \tag{23}$$

where $\Delta\lambda_{\text{tree},X} \equiv 2\lambda_{HX}^2 v_X^2 / m_X^2$ for $X = S, \phi$.[8] This means that if one of the contributions is suppressed, so will the other be, despite the difference in the sizes of the associated VEVs and portal couplings. Thus, in this study, we will not be able to proceed in the same manner as in [25] and completely rely on sizable tree-level shifts in the Higgs self-coupling to render the Higgs potential fully stable. As will be shown below, one-loop effects will be essential to achieve stabilization.

[6]Note that the condition $\mu \ll M_N$ is not necessary to obtain Eq. (18). However, we use it in Appendix A to obtain the approximate analytical expression for the (block-)diagonalization matrix, given that this limit of small lepton number violation is of particular interest in this work.

[7]In [7], $\lambda_{HS} \sim 10^{-12}$, which turns out to be the main reason for their shift in the Higgs quartic $\lambda$ being negligible.

[8]Note that the Higgs self-coupling does not contribute to this tree-level matching since the Higgs boson remains in the symmetric phase at $\Lambda_{GW}$. Its one-loop contributions are included in the RG running below this scale.

#### 2. Fermionic sector

The field-dependent masses for the RHNs can be obtained by diagonalizing the $7\times 7$ matrix in Eq. (17). Such diagonalization can be carried out analytically through a series expansion, as in Ref. [26]. For simplicity, we consider the case where $\mu$ is a complex-diagonal $2\times 2$ matrix, as in Ref. [27]. Here, we show how to obtain the first-order terms, independent of $h$, and we leave the more involved computation of the field-dependent parts to Appendix C.

As a first step, the pairs $(N_1, \xi_1)$ and $(N_2, \xi_2)$ can be decoupled by a reordering of rows and columns of the mass matrix, defined as $r_2 \leftrightarrow r_3, c_2 \leftrightarrow c_3$. This is then followed by an almost-maximal rotation of the heavy $4\times 4$ block (lower-right $4\times 4$ block), achieved via the rotation matrix

$$U = \begin{pmatrix} \mathbb{I}_{3\times 3} & 0 & 0 \\ 0 & U_1 & 0 \\ 0 & 0 & U_2 \end{pmatrix}, \tag{24}$$

where

$$U_i \equiv \begin{pmatrix} -i\cos\alpha_i e^{i\theta_i/2} & \sin\alpha_i e^{i\theta_i/2} \\ i\sin\alpha_i e^{-i\theta_i/2} & \cos\alpha_i e^{-i\theta_i/2} \end{pmatrix}, \tag{25}$$

with $i = 1,\ 2$ for heavy neutrino flavors. Above, $e^{i\theta_i}$ denotes the phases of $\mu_{ii}$, while $\sin\alpha_i \approx \frac{1}{\sqrt{2}}(1-\frac{\varepsilon_i}{4})$ and $\cos\alpha_i \approx \frac{1}{\sqrt{2}}(1+\frac{\varepsilon_i}{4})$, with $\varepsilon_i \equiv \mu_{ii}/M_{N_i} \ll 1$. More specifically,

$$\tilde{M}_\nu(h) = \begin{pmatrix} (0)_{3\times 3} & \tilde{M}_{D,1}(h) & \tilde{M}_{D,2}(h) \\ \tilde{M}^T_{D,1}(h) & M_N^{(1)} & 0 \\ \tilde{M}^T_{D,2}(h) & 0 & M_N^{(2)} \end{pmatrix}, \tag{26}$$

with the lower diagonal blocks given by

$$M_N^{(i)} = \mathrm{diag}\left[M_{N_i}\left(1-\frac{\varepsilon_i}{2}\right), M_{N_i}\left(1+\frac{\varepsilon_i}{2}\right)\right], \tag{27}$$

and the new Yukawa matrices defined as $\tilde{M}_{D,i}(h) \equiv y^{\pm}_{\alpha i} h/\sqrt{2}$. The matrices $y^{\pm}_{\alpha i}$ ($\alpha = 1,2,3$ and $i = 1,2$) are related to the original Yukawa matrix by the following relations:

$$y^{\pm}_{\alpha i} \approx \frac{i e^{-i\theta_i/2}}{\sqrt{2}}\left(1 \pm \frac{\varepsilon_i}{4}\right)(y_\nu)_{\alpha i}. \tag{28}$$

Performing a second series expansion, this time centered around a zero rotation angle, leads to the following fermionic field-dependent masses up to order $h^2$:

$$M^{\pm}_{N_i}(h) \approx M_{N_i}\left(1 \pm \frac{\varepsilon_i}{2}\right) + \frac{1}{2M_{N_i}}\sum_{\alpha=1}^{3} |y^{\mp}_{\alpha i}|^2 h^2. \tag{29}$$

Above, we ignored terms $\mathcal{O}(h^4)$ under the assumption that shifts to the Higgs self-coupling coming from the fermion sector are negligible (see below). With these fermionic field-dependent masses, Eq. (B6) can be used to compute the associated threshold corrections to the Higgs mass parameter, which reads as

$$\Delta m_i^2 \approx \frac{1}{16\pi^2} M^2_{N_i} \sum_\alpha |y_{\alpha i}|^2 \left(1 + 2\log\frac{M^2_{N_i}}{Q^2}\right), \tag{30}$$

where we neglected terms proportional to $\varepsilon_i$ assuming $\varepsilon_i \ll 1$.

## III. NUMERICAL ANALYSIS

In this section, we perform a numerical analysis of our minimal model described in the previous section.

### A. Experimental bounds

This subsection summarizes the observables that we take into account to constrain the model's parameter space.

#### 1. Light neutrino masses and IR parameters

We employ the Casas-Ibarra parametrization for light SM neutrino masses, presented in Appendix A. In our model, the light neutrino masses are generated through the inverse seesaw and lie within the experimental bounds shown in Table V. However, this is achieved at the GW scale and requires considering the RGE effects on the couplings down to lower energies. Moreover, it will be required that the correct one-loop Higgs mass and Higgs self-coupling in the $\overline{\mathrm{MS}}$ scheme, and within the convention of Eq. (21), are reproduced at the top mass scale [28], i.e.,

$$m_H^2(m_t) = (93.5)^2\ \mathrm{GeV}^2, \qquad \lambda(m_t) = 0.128. \tag{31}$$

Since the aim of this analysis is proof of concept rather than precision calculations, when scanning the parameter space of the model, a tolerance will be allowed for the different physical quantities, which do not necessarily lie within the current experimental bounds. For instance, even though $m_H^2$ has been measured to an accuracy of $\sim 0.1\%$ [29], the analysis below will consider tolerances of $\mathcal{O}(10\%)$.

#### 2. Lepton flavor violating processes

LFV We take into account the branching ratio of lepton flavor violating (LFV) processes, which are mediated by right-handed neutrinos $N$ (hereinafter we drop the subscript $R$ from RHNs for brevity) at the one-loop level, as shown in the following Feynman diagrams:

where $\alpha, \beta = e, \mu, \tau$ correspond to the different lepton flavors, and $i$ enumerates the different RHNs. The current experimental bounds on the branching ratios of the considered LFV processes are summarized in Table II.

The contribution from the RHNs to these LFV branching ratios is given by, see, e.g., [24,32,33],

$$\mathrm{BR}(\ell_\alpha \to \ell_\beta \gamma) = \frac{\alpha^3 \sin^2\theta_W}{256\pi^2} \left(\frac{M_{\ell_\alpha}}{M_W}\right)^4 \frac{M_{\ell_\alpha}}{\Gamma_{\ell_\alpha}} |G_{\alpha\beta}|^2, \tag{32}$$

where $\alpha = e^2/4\pi$ is the fine-structure constant, $\theta_W$ is the Weinberg angle, $M_{\ell_\alpha}$ is the mass of the decaying lepton, $M_W$ is the $W$ boson mass, $\Gamma_{\ell_\alpha}$ is the total decay width of the decaying lepton, and

$$G_{\alpha\beta} \equiv \sum_j U^*_{\alpha j} U_{\beta j} G_\gamma\left(\frac{M^2_{N_j}}{M^2_W}\right). \tag{33}$$

Above $U$ is the unitary matrix that diagonalizes the neutrino mass matrix $M_\nu$ (17), i.e.,

$$U^T M_\nu U = \mathrm{diag}(m_{\nu_1}, m_{\nu_2}, m_{\nu_3}, M_{N_1}, M_{N_2}, M_{N_3}, M_{N_4}). \tag{34}$$

In Eq. (33), $G_\gamma(M_N^2/M_W^2)$ is a photonic composite form factor obtained from expanding the loop integrals up to the first nonvanishing order [34], and is given by

$$G_\gamma(x) = -\frac{2x^3 + 5x^2 - x}{4(1-x)^3} - \frac{3x^3}{2(1-x)^4} \ln x. \tag{35}$$

TABLE II. Experimental upper bounds on the branching ratios (BRs) of the LFV processes $\mu \to e\gamma$, $\tau \to e\gamma$, and $\tau \to \mu\gamma$.

| Branching ratio | Exp. bound |
|---|---|
| $\mathrm{BR}(\mu \to e\gamma)$ | $< 4.2 \times 10^{-13}$ [30] |
| $\mathrm{BR}(\tau \to e\gamma)$ | $< 1.5 \times 10^{-10}$ [31] |
| $\mathrm{BR}(\tau \to \mu\gamma)$ | $< 1.5 \times 10^{-10}$ [31] |

Furthermore, the total decay width of the muon is in agreement with the experimental data, and its analytic form is [32]

$$\Gamma_\mu = \frac{G_F^2 M_\mu^5}{192\pi^3}\left(1 - 8\frac{M_e^2}{M_\mu^2}\right)\left[1 + \frac{\alpha}{2\pi}\left(\frac{25}{4} - \pi^2\right)\right], \tag{36}$$

where $G_F$ is the Fermi constant, $M_\mu$ is the muon mass, and $M_e$ is the electron mass. On the other hand, the total decay width of the tau lepton is measured to be $\Gamma_\tau = (2.267 \pm 0.004) \times 10^{-12}$ GeV [35].

#### 3. Heavy neutral lepton mixing

Current experiments, including beam dump and collider experiments, as well as astrophysical and cosmological observations, are sensitive to the mixing of heavy neutral leptons or RHNs with the SM particles that have masses approximately below the electroweak scale. For instance, in Ref. [12], the most relevant bounds for RHN masses in the range $\mathcal{O}(1)$ GeV $< M_N < \mathcal{O}(10^2)$ GeV are collected and displayed at the 90% confidence level. These bounds on the entries of the neutrino mixing matrix are derived under the assumption of single-flavor dominance, where the singlet heavy neutral lepton couples predominantly to one particular lepton flavor. Even though this is not an underlying assumption of our model, the expectation is that if such an assumption is not made, the bounds will become less stringent. In the following numerical analysis, we take such constraints into account.

#### 4. Scalar mixing

Further constraints on the model's parameter space could be achieved by requiring that the mixings between our BSM scalars and the SM Higgs boson are in agreement with current constraints from Higgs couplings and BSM scalar searches. If $\theta_m$ is the mixing angle between the Higgs boson and a BSM scalar, a model-independent limit at 95% CL is given by [36]

$$|\sin\theta_m| \lesssim 0.2. \tag{37}$$

TABLE III. Conditions that must be satisfied for the model to be UV complete and conformal, capable of producing the correct light neutrino masses, a radiatively generated Higgs mass, and a fully stable scalar potential.

| | |
|---|---|
| Inverse seesaw | $\mu \ll M_D, M_N$ |
| SM neutrino masses | $m_\nu \approx -M_D (M_N^T)^{-1} \mu M_N^{-1} M_D^T \overset{!}{=} m_\nu^{\mathrm{exp}}$ |
| Higgs mass $m_H^2$ | $\Delta m^2_{N,S,\phi} \le m_H^2/\Delta_{\mathrm{FT}}$ |
| Vacuum stability | $\lvert\Delta\lambda_{N_R}\rvert < \mathcal{O}(10^{-2}), \mathcal{O}(10^{-3}) < \Delta\lambda_\phi < \mathcal{O}(10^{-1})$ |
| Stable $V_{\mathrm{eff}}$ | $B > 0$ |
| Perturbativity | $\mathcal{P} < 4\pi \;\; \forall \;\; \Lambda < M_{\mathrm{Pl}}$ |
| No extra flat directions | GW conditions $= \{\varnothing\}$ if $\Lambda_{\mathrm{GW}} < \Lambda < M_{\mathrm{Pl}}$ |

Moreover, the presence of BSM scalars affects all couplings of the physical Higgs boson. For instance, the trilinear coupling $g_{hZZ}$ between the physical Higgs scalar $h$ and two $Z$ gauge bosons in the BSM model gets modified to the one in the SM ($g_{hZZ}^{\mathrm{SM}}$) by [37]

$$\delta g_{hZZ} = \frac{g_{hZZ}}{g_{hZZ}^{\mathrm{SM}}} - 1 = \cos\theta_m - 1. \tag{38}$$

Experimental bounds on this trilinear coupling are given by the ATLAS and CMS Collaborations and can also result in further constraining the model's parameter space. An analogous situation holds for the trilinear and quartic couplings of the Higgs boson, which also get modified when introducing BSM scalars. However, they are significantly less constrained than the $hZZ$ coupling. A detailed phenomenological study of the scalar sector of this model is beyond the scope of this paper; therefore, we ignore such constraints in the following numerical analysis, but we refer the reader to [14,37] for related studies.

### B. Parameter space for the minimal model

After applying the GW conditions (7), there are at least ten new physics parameters.[9] For the numerical analysis, each set of boundary conditions defined at the scale $\Lambda_{\mathrm{GW}}$ is checked to determine whether the correct low-energy observables are reproduced and to ensure that the model remains classically conformal, perturbative, and stable up to the Planck scale. Table III summarizes the consistency conditions that must be satisfied. These constraints impose bounds on the various couplings and masses associated with the scalars $S$ and $\phi$, which we briefly summarize below. We then use the scan points that satisfy these conditions as boundary values for the RG running required to obtain the low-energy observables. Each row in Table III, except for the last two, corresponds to constraints that can be enforced at $\Lambda_{\mathrm{GW}}$, the scale at which the boundary conditions for the RG flows are defined. In the following, we comment on each of these conditions.

[9]The actual number depends on the assumptions made. For instance, assuming $M_N$ to be diagonal reduces the number of associated parameters by half.

The condition $\mu \ll M_D, M_N$ is a general assumption in models based on the inverse seesaw mechanism. Since the light neutrino masses are obtained by diagonalizing the full neutrino mass matrix, including the heavy neutrinos, the first-order result is only valid when such conditions hold. The second condition ensures that the correct light neutrino masses are achieved. As mentioned above, this condition is imposed at $\Lambda_{\mathrm{GW}}$, but $m_\nu$ runs when considering lower energy scales. Therefore, this constraint is only valid if the running is small enough so that $m_\nu$ remains within the $3\sigma$ experimental bounds. As will be shown below, this will generally be the case. The third condition ensures that the shift in the Higgs mass induced by the new physics is of the correct order of magnitude, up to some level of fine-tuning parametrized by $\Delta_{\mathrm{FT}}$. Equations (B8) and (30) show that these shifts are negative/positive for scalars/fermions. With this in mind, a fine-tuning parameter $\Delta_{\mathrm{FT}}$ is defined as

$$\Delta_{\mathrm{FT}} \equiv \left( \sum_i \frac{|\Delta m_i^2|}{m_H^2} \right)^{-1}, \tag{39}$$

where $\Delta m_i^2$ are the contributions to the Higgs mass squared $m_H^2$, with $i$ running over the appropriate BSM particles. For instance, if the $\Delta m^2$ threshold corrections induced by the heavy scalars are of $\mathcal{O}(1)$ TeV$^2$, the associated boundary conditions are assigned a fine-tuning of $\Delta_{\mathrm{FT}} \sim 1\%$. The following analysis focuses on the range $\Delta_{\mathrm{FT}} \in [0.1, 100]\%$, where $\Delta_{\mathrm{FT}} = 100\%$ means no fine-tuning. Setting this requirement on the level of fine-tuning results in an upper bound on the threshold corrections to the Higgs mass and excludes solutions that need fine-tuning at a precision smaller than 0.1%.

It is well known that the SM Higgs self-coupling $\lambda$ runs to negative values in the UV [9–11]. More specifically, at one-loop level in the $\overline{\mathrm{MS}}$ scheme,

$$\lambda(\Lambda \gtrsim 10^6\ \mathrm{GeV}) < 0 \sim -\mathcal{O}(10^{-2}). \tag{40}$$

This relation indicates that the BSM particles must induce a positive shift in $\lambda$ of order $\mathcal{O}(10^{-2})$ to render the Higgs scalar potential fully stable. A positive shift in the Higgs quartic $\lambda$ can be produced through scalars, as shown in Eq. (B9), and see, e.g., [25,38]. This justifies the constraints in the fourth row of Table III.[10] Furthermore, we note that in the context of the Gildener-Weinberg framework [20], for the radiatively generated extremum along the flat direction to be a minimum, the $B$ function, defined in Eq. (B3), must be positive. This corresponds to the fifth constraint in Table III.

The last two constraints in Table III become relevant once RGEs are included in the analysis. Specifically, after selecting the set of boundary conditions for which the low-energy physics constraints are satisfied, these are used as initial values for a bottom-up RG running from the GW scale up to the Planck scale. In doing so, we check at each RGE step that the model remains perturbative, stable, and classically conformal. Perturbativity is satisfied if no coupling grows larger than $4\pi$ before reaching the Planck scale. Stability requires that the scalar quartic couplings remain positive, while classical conformality is achieved if no further flat directions are developed above the GW scale. Such extra flat directions would break classical conformal symmetry before the GW scale, rendering the GW analysis invalid. In practice, it must be checked that the conditions derived in Appendix B are not satisfied above the GW scale.

In summary, we adopt a version of the Casas-Ibarra parametrization,[11] use the GW conditions (7) at $\Lambda_{GW}$, fix the mixing angles to their measured central values (see Table V), and assume, without loss of generality, $M_N$ to be real diagonal and $\mu$, for simplicity, to be complex diagonal. With this, the model's parameter space at the GW scale $\Lambda_{\mathrm{GW}}$ can be described by the following set of parameters:

$$\mathcal{P} = \{v_S, \alpha, y_N^{1,2}, y_M^{1,2}, \theta_{1,2}, \rho, \beta, \lambda_{HS}, \lambda_{H\phi}, \lambda_S\}. \quad (41)$$

[12] In the following subsections, we derive explicit constraints on these model parameters.

### 1. Boundary conditions and constraints

In this subsection, we employ the conditions from Table III to derive constraints for some of the couplings in the parameter space $\mathcal{P}$.[13] Referring to Eq. (B3) and requiring that $B > 0$ at $\Lambda_{\mathrm{GW}}$, we arrive at bounds for the $\phi$ and $H$ scalar masses given by

$$(M_\phi^0)^4 > 8M_{N_2}^4 - 2|\Delta m_{\mathrm{tree}}^2|^2, \quad (42)$$

$$(M_H^0)^2 = \lambda_{HS} v_S^2 + \lambda_{H\phi} v_\phi^2 = |\Delta m_{\mathrm{tree}}^2| \overset{!}{<} m_H^2/\Delta_{\mathrm{FT}}. \quad (43)$$

Above, we have assumed that the heaviest, almost degenerate pair of RHNs with mass $M_{N_2}$ is significantly heavier than the lightest pair with mass $M_{N_1}$, i.e., $M_{N_2} \gg M_{N_1}$. Also, in Eq. (43), we have made explicit the required suppression of the initial shift in $m_H^2$ for a given value of the fine-tuning parameter $\Delta_{\mathrm{FT}}$. The leading-order scalar field masses for $\phi$ and $S$ at $\Lambda_{\mathrm{GW}}$, obtained from Eq. (14), are

$$(M_\phi^0)^2 = \frac{8\lambda_S}{\sin^2\alpha} v_S^2, \qquad (M_S^0)^2 = 8B(v_S^2 + v_\phi^2). \quad (44)$$

We now consider conditions involving the $\Delta m^2$ threshold corrections and focus on the nonlogarithmic terms. The constraint (43) leads to a suppression of the portal couplings as

$$\lambda_{HS} < \frac{m_H^2}{v_S^2 \Delta_{\mathrm{FT}}}, \qquad \lambda_{H\phi} < \frac{m_H^2}{v_S^2 \tan^2\alpha \Delta_{\mathrm{FT}}}. \quad (45)$$

These relations show that choosing the angle $\alpha$ small allows us to relax the upper bound on the portal coupling $\lambda_{H\phi}$ relative to the one on $\lambda_{HS}$. This is a major difference with respect to [7], where $\alpha = 0$ is chosen and, therefore, the only portal coupling that contributes to $\Delta\lambda$ at tree level is $\lambda_{HS}$. This portal is highly suppressed by their analog of Eq. (45), leaving tree-level contributions to $\Delta\lambda$ ineffective. Moreover, the fact that one-loop contributions to $\Delta m^2$ involve a product of the portal coupling with the associated BSM scalar mass [see Eq. (B8)] renders any effect on the RG running of $\lambda$ negligible at one-loop order when low fine-tuning and large scalar masses are required. More precisely, in [7] the RHN masses must lie in the range 10–500 PeV, and the scalar with vanishing VEV must be more massive than these RHNs to achieve a stable one-loop effective potential. This results in highly suppressed shifts in $\lambda$, leaving no room in the parameter space where the Higgs potential may be fully stabilized.

Coming back to the model under study, assuming $\alpha \ll 1$ it is possible to write the scalar one-loop contributions [see Eq. (B8), where $\kappa_{S/\phi} \sim \lambda_{HS/\phi}$] as

$$|\Delta m_{S,\mathrm{loop}}^2| \sim \frac{\lambda_{HS}}{16\pi^2} (M_S^0)^2 < m_H^2/\Delta_{\mathrm{FT}},$$

$$|\Delta m_{\phi,\mathrm{loop}}^2| \sim \frac{\lambda_{H\phi}}{16\pi^2} (M_\phi^0)^2 < m_H^2/\Delta_{\mathrm{FT}}. \quad (46)$$

On the other hand, Eq. (B9) implies that $\Delta\lambda$ shifts are also proportional to these portal couplings, meaning that sizable shifts in the Higgs self-coupling will be possible for large enough portals. Combining this requirement with Eq. (46) results in an upper bound on the mass of the heaviest scalar $\phi$. This is the first hint toward what will turn out to be the strongest source of tension in this model, requiring that $B > 0$ at $\Lambda_{\rm GW}$ pushes the BSM scalar masses to higher values as the RHN masses are increased [see Eq. (43)]. However, requiring sizable shifts to the Higgs self-coupling with a low level of fine-tuning pushes the scalar masses to lower values. One cannot freely alleviate this tension by taking smaller RHN masses. First, the RHN masses should not be much smaller than $M_N \sim 1$ GeV so that the singlet neutrinos can decay before big bang nucleosynthesis [39,40]. Second, large enough RHN masses are needed to induce the correct Higgs mass at low energies.

[10]Note that even though the focus is on the Higgs self-coupling, ensuring vacuum stability also requires the other quartic couplings in the theory to remain positive, and the portal couplings cannot be too negative [37].

[11]See Appendix A for details of the parametrization and the definition of the associated parameters $\rho$ and $\beta$.

[12]Note that the Higgs self-coupling, which is in principle a free parameter, is not included here. The reason for this is that in the analysis below, we will only be interested in RG solutions reproducing the correct value of $\lambda$ in the IR.

[13]Note that all couplings, and consequently also masses, are evaluated at $\Lambda_{\mathrm{GW}}$ where the GW conditions hold.

Note that if the first inequality in Eq. (46) is satisfied, so is the second. This follows from choosing $\alpha \ll 1$, where $\lambda_{HS}$ is more suppressed than $\lambda_{H\phi}$ due to Eq. (45), and the scalon mass is suppressed relative to the other scalar by at least a loop factor [see Eq. (B5)]. These upper bounds can be tightened by taking into account the perturbativity requirement

$$\lambda_\phi \overset{@\Lambda_{\rm GW}}{=} \lambda_S/\tan^4\alpha \sim \lambda_S/\alpha^4 \overset{!}{<} 1. \tag{47}$$

With this condition, Eq. (46) implies

$$|\Delta m^2_{\phi,\rm loop}| < 0.1 m_H^2/\Delta_{\rm FT}. \tag{48}$$

This new upper bound shows that for a given fine-tuning parameter $\Delta_{\rm FT}$, the one-loop contribution to $\Delta m^2_\phi$ is about 1 order of magnitude more suppressed than the tree-level contribution. Looking at Eq. (46), this results in even more stringent upper bounds on the scalar masses.

As mentioned above [Eq. (23)], the tree-level corrections to $\lambda$ are equally suppressed for $S$ and $\phi$ due to the GW conditions. From Eq. (22), it can be seen that an upper bound for $\Delta\lambda_{\rm tree}$ stems from a lower bound for $\lambda_S$. Since $(M^0_\phi)^2 \propto \lambda_S$ [Eq. (44)], a lower bound for $M^0_\phi$ translates to a lower bound on $\lambda_S$, which can be obtained from the stability condition of the effective potential.

As anticipated above, the effect of having $\lambda_{H\phi}$ sizable is to make $\phi$, which is the heaviest particle in the model, lighter. This can be seen from Eq. (46) with $\kappa_\phi \approx \lambda_{H\phi} \gtrsim \mathcal{O}(10^{-1})$, which leads to the following upper bound on the mass of $\phi$:

$$(M^0_\phi)^2 < 0.01 m_H^2/\Delta_{\rm FT}. \tag{49}$$

Also, putting Eqs. (42) and (49) together leads to the following upper bound on the RHN masses:

$$M_N < 0.05 m_H/\sqrt{\Delta_{\rm FT}}. \tag{50}$$

Therefore, by requiring low fine-tuning, i.e., $\Delta_{\rm FT} \gtrsim 0.1\%$, and the shifts in $\lambda$ to be dominated by the one-loop contributions, all the new-physics particles must be at or below the TeV scale.[14]

We find that in the range of interest where $\Lambda_{\rm GW} > \mathcal{O}(10^3)$ GeV,[15] the GW scale is usually of the same order of magnitude as the heaviest scalar's mass $M_\phi$. As seen above, this mass is suppressed when we require large portals. Thus, obtaining larger GW scales with sizable portals is only possible with increasingly larger levels of fine-tuning. On the contrary, without the condition $\Delta\lambda > \mathcal{O}(10^{-3})$, the GW scale can take values in the range of interest regardless of the level of fine-tuning.

The main conclusion to be drawn from this section is that sizable shifts in $\lambda$, both at tree or one-loop level, favor smaller RHN masses. How small these need to be depends on the level of fine-tuning allowed. However, smaller RHN masses must be accompanied by larger Yukawa couplings so that the positive contribution of the RHNs to the shifts in the Higgs mass, as in Eq. (30), is of the correct order. Furthermore, sizable shifts in the Higgs self-coupling are only possible when the heaviest scalar has a mass around or below the TeV range in the case of low fine-tuning. A closer look into these sources of tension and their implications will be taken below once the RG runnings are discussed.

#### *2. IR and UV consistency conditions*

After selecting a set of boundary conditions from the input scan discussed above, the following step corresponds to integrating out the heaviest particle(s) of the model. This will induce an initial shift in the Higgs mass parameter, which is taken to be zero above the conformal symmetry-breaking scale, and in the Higgs self-coupling. Then, the model's $\beta$ functions are used to run down all parameters of the effective theory to the top mass scale $m_t$ where the low-energy checks are performed.[16] The first condition consists of reproducing the correct Higgs mass with a given tolerance,[17] i.e.,

$$0.75(m_H^2(m_t))_{\rm exp} < m_H^2(m_t) < 1.25(m_H^2(m_t))_{\rm exp}, \tag{51}$$

where $(m_H^2(m_t))_{\rm exp}$ stands for the measured central value of the Higgs mass parameter at the top mass scale, and the assigned tolerance is 25%. Similarly, the induced Higgs self-coupling is also required to lie around its central value as measured in the experiment with 10% tolerance, i.e.,

$$0.9(\lambda(m_t))_{\mathrm{exp}} < \lambda(m_t) < 1.1(\lambda(m_t))_{\mathrm{exp}}. \tag{52}$$

[14]Note that the analysis in [7] was not concerned with sizable shifts in $\lambda$, so small portal couplings were allowed, and large scalar masses were easily attainable.

[15]A conformal model with a low-lying SBCS scale is much more likely to run into problems when trying to preserve perturbativity and conformality in the UV.

[16]The associated one-loop $\beta$ functions at scales below and above the conformal breaking scale were obtained with PyR@TE3 [17] and can be found in the appendix of Ref. [41].

[17]As mentioned above, the focus of this study is proof of concept rather than precision measurements. Thus, the tolerance considered here is less restrictive than the current experimental uncertainty.

TABLE IV. Parameter ranges for a first parameter space scan aiming to show the viability of our UV complete conformal model based on the IS(2,2) scenario described above.

| $v_S$ | $\alpha$ | $\lambda_S$ | $\lambda_{H\phi}$ | $\lambda_{HS}$ | $y_N$ | $\rho$ | $y_M$ |
|---|---|---|---|---|---|---|---|
| $10^{[6,10]}$ GeV | $10^{[-5,0]}$ | $10^{[-20,0]}$ | $10^{[0,7]}$ $\mathrm{GeV}^2/v_\phi^2$ | $10^{[0,7]}$ $\mathrm{GeV}^2/v_S^2$ | $10^{[0,6]}$ $\mathrm{GeV}/v_S$ | $10^{[-4,0]}$ | $10^{[-10,0]}$ |

Even though the adapted Casas-Ibarra parametrization detailed in Appendix A fixes the light neutrino masses to the experimentally measured values,[18] this is done at the scale of conformal symmetry breaking. One must verify that the RG running does not take these values outside the experimental bounds after running down to low energies. In other words, we require that $m_\nu$ satisfies the experimental $3\sigma$ constraint specified in Table V.

All boundary conditions satisfying these low-energy constraints are selected to constitute a filtered version of the original input set. This new set is used as input for a bottom-up RG running starting at the conformal symmetry breaking scale $\Lambda_{\mathrm{GW}}$ and ending at the Planck scale $M_{\mathrm{Pl}}$. After solving the RGEs numerically, it must be checked that at each step of the RG running the couplings remain perturbative, i.e.,

$$\mathcal{P}(\Lambda_{\mathrm{GW}} < \Lambda < M_{\mathrm{Pl}}) < 4\pi, \tag{53}$$

where $\mathcal{P}$ stands for all couplings parametrizing the model's parameter space. Moreover, for the model to remain classically conformal above the GW scale, no extra flat directions can develop. To make sure that this is the case, it must be verified that none of the GW relations derived in Appendix B are satisfied. More specifically, one must check, at each RG step, that the following equations are not satisfied:

$$\begin{aligned} \lambda = \lambda_S = \lambda_\phi &= 0, \\ \lambda_{HS}^2 - 4\lambda\lambda_S = \lambda_{H\phi}^2 - 4\lambda\lambda_\phi = \lambda_{S\phi}^2 - 4\lambda_S\lambda_\phi &= 0, \\ \lambda_{HS}^2\lambda_\phi + \lambda_{H\phi}^2\lambda_S + \lambda_{S\phi}^2\lambda - \lambda_{HS}\lambda_{H\phi}\lambda_{S\phi} - 4\lambda\lambda_S\lambda_\phi &= 0. \end{aligned} \tag{54}$$

Finally, a fully stabilized Higgs potential is achieved by ensuring that the Higgs self-coupling remains positive along all energy scales up to the Planck scale, i.e.,

$$\lambda(\Lambda_{\mathrm{GW}} < \Lambda < M_{\mathrm{Pl}}) > 0. \tag{55}$$

Even though one could, in principle, relax this condition such that the Higgs self-coupling becomes negative along a short energy range, the fact that $\lambda$ changes sign implies that classical scale symmetry will be broken at a scale higher than the previously stipulated $\Lambda_{\mathrm{GW}}$.

### 3. UV scale invariance

This subsection aims to show that our classically conformal realization of the neutrino option based on an inverse seesaw framework is viable, with RHN masses in the range $1 - 10^6$ GeV. To this end, a linear-log scan is carried out over the parameter ranges shown in Table IV.

In particular, the scalon VEV $v_S$ is chosen to lie in the range $10^6$–$10^{10}$ GeV because random choices of parameter values will most frequently result in highly suppressed shifts in the Higgs self-coupling. Therefore, in most cases, $\lambda$ will be expected to resemble the SM RG running, i.e., it will run negative around $\Lambda \sim 10^6$ GeV at one-loop order. Since the conformal breaking scale generally satisfies $\Lambda_{\mathrm{GW}} \lesssim v_S$ [see Eq. (B4)] and $\lambda = 0$ is one of the GW relations in Eq. (54) which would lead to an extra flat direction, the model usually cannot remain classically scale invariant in the UV if $\Lambda_{\mathrm{GW}} \lesssim 10^6$ GeV. Also, the smallness of $\lambda_S$ comes from the GW relation $\lambda_\phi = \lambda_S \tan^{-4}\alpha$ combined with the perturbativity condition $\lambda_\phi < 1$. The portals $\lambda_{HS}$ and $\lambda_{H\phi}$ are chosen to acquire the largest possible values that would still satisfy the allowed level of fine-tuning, which here is taken to be $\Delta_{\mathrm{FT}} \gtrsim 0.1\%$. The couplings $y_N$ are chosen such that the RHN masses lie in the range $1 - 10^6$ GeV, while $\rho$ and $y_M$, which are found not to play a major role in this particular analysis, are well represented by the range detailed above.

The induced Higgs mass parameter and Higgs self-coupling after a top-down RG running from the GW scale to the top mass scale are computed for different UV boundary conditions. We find that boundary conditions leading to an initial shift in $m_H^2$ of a similar order as the low-energy Higgs mass parameter, but with an opposite sign, are favored. In other words, solutions with low fine-tuning as defined in Eq. (39) are more likely to result in the correct radiatively induced Higgs mass parameter. Also, many boundary conditions lead to RG solutions where $\lambda$ grows too large at low energies, escaping the allowed range, and this occurs mainly when $m_H^2$ runs too large as well. In other words, for lower values of $m_H^2(m_t)$, $\lambda(m_t)$ tends to reside in the correct ballpark, and it is only for $m_H^2(m_t) \gtrsim m_H^2(m_t)_{\rm exp}$ where the solutions start to become highly unstable.

[18]Recall that even though the absolute values of the light neutrino masses have not been measured, in the IS(2,2) scenario one of the light neutrinos must be massless, and therefore the two measured shifts $\sqrt{|\Delta m^2_{\mathrm{sol}}|}$ and $\sqrt{|\Delta m^2_{\mathrm{atm}}|}$ become the absolute masses of the other two neutrinos.

The numerical evaluation also corroborates that, since most solutions will resemble the quartic RG running of the SM, the associated GW scales must be larger than the scale at which the SM Higgs self-coupling changes sign. If not, an extra flat direction develops at a scale larger than the originally postulated GW scale, rendering the analysis invalid.

As mentioned above, the light neutrino masses are fixed to the experimental central values at the conformal breaking scale, and they are subject to RG running when going to lower scales. We found that light neutrino masses stemming from practically all solutions satisfying the low-energy requirements remain inside the experimental bounds after the top-down RG running from the GW scale to the top mass scale. This situation is even better realized for the subset of solutions that also remain classically conformal and perturbative up to the Planck scale. In other words, the RG running of the light neutrino masses is not large enough to take them out of the experimental bounds.

In Sec. III A, we discussed how the introduction of RHNs in the SM can influence the branching ratios of LFV processes. Experimental bounds on the branching ratios of the processes $\mu \to e + \gamma$, $\tau \to e + \gamma$, and $\tau \to \mu + \gamma$ exist and are detailed in Table II. We will refrain from showing the results of scans in the current scenario and will instead discuss them later in the context of the next-to-minimal model.

#### *4. Vacuum stability*

In this subsection, we explore the possibility of stabilizing the Higgs scalar potential in the conformal realization of the neutrino option. For this, we use boundary conditions that satisfy $\Delta\lambda > \mathcal{O}(10^{-3})$ with $\Delta\lambda$ given by Eqs. (B9) and (22). Also, we focus on values of the portal $\lambda_{H\phi}$ in the range $\mathcal{O}(10^{-2}) < \lambda_{H\phi} < 1$.

Figure 1 shows that in this context, inducing the correct Higgs mass at the top mass scale restricts the RHN masses to lie in the range $\sim(0.3, 3)$ TeV. Any RHN mass below this range results in a Higgs mass below the experimental value, and the opposite occurs for RHN masses above this range. The larger the initial shift on $m_H^2$ induced by integrating out $\phi$, the heavier the RHNs need to be for the RG running to take the initial negative value to the correct positive one at lower scales.

Even though small shifts in the Higgs self-coupling are possible, requiring $\Delta\lambda$ to be large enough to render the Higgs potential fully stable inevitably results in severe instabilities. The source of such instabilities lies in sizable shifts in the Higgs self-coupling being connected to large Yukawas or, equivalently, large $\beta$ functions at the GW scale. As shown in the left panel of Fig. 1, sizable $\Delta\lambda$ shifts are attainable for larger RHN masses only at the expense of introducing some level of fine-tuning ($\Delta_{\rm FT} < 30\%$). Even then, if these masses become too large, the RG running results in a Higgs mass above the experimental value. Thus, to keep the level of fine-tuning as low as possible, RHN masses below the TeV scale are required. However, choosing small enough RHN masses to achieve sizable $\Delta\lambda$ shifts at tree level will then require the Yukawa couplings $Y_\nu$ to be large enough so that the RG running of the Higgs mass is enhanced and $m_H^2(m_t)$ is not below the experimental value. Finally, the reason why this results in severe instabilities is that large Yukawas also enhance the RG running of the Higgs self-coupling, since $\beta_\lambda \sim Y_\nu^4$, in a way that completely destabilizes it.

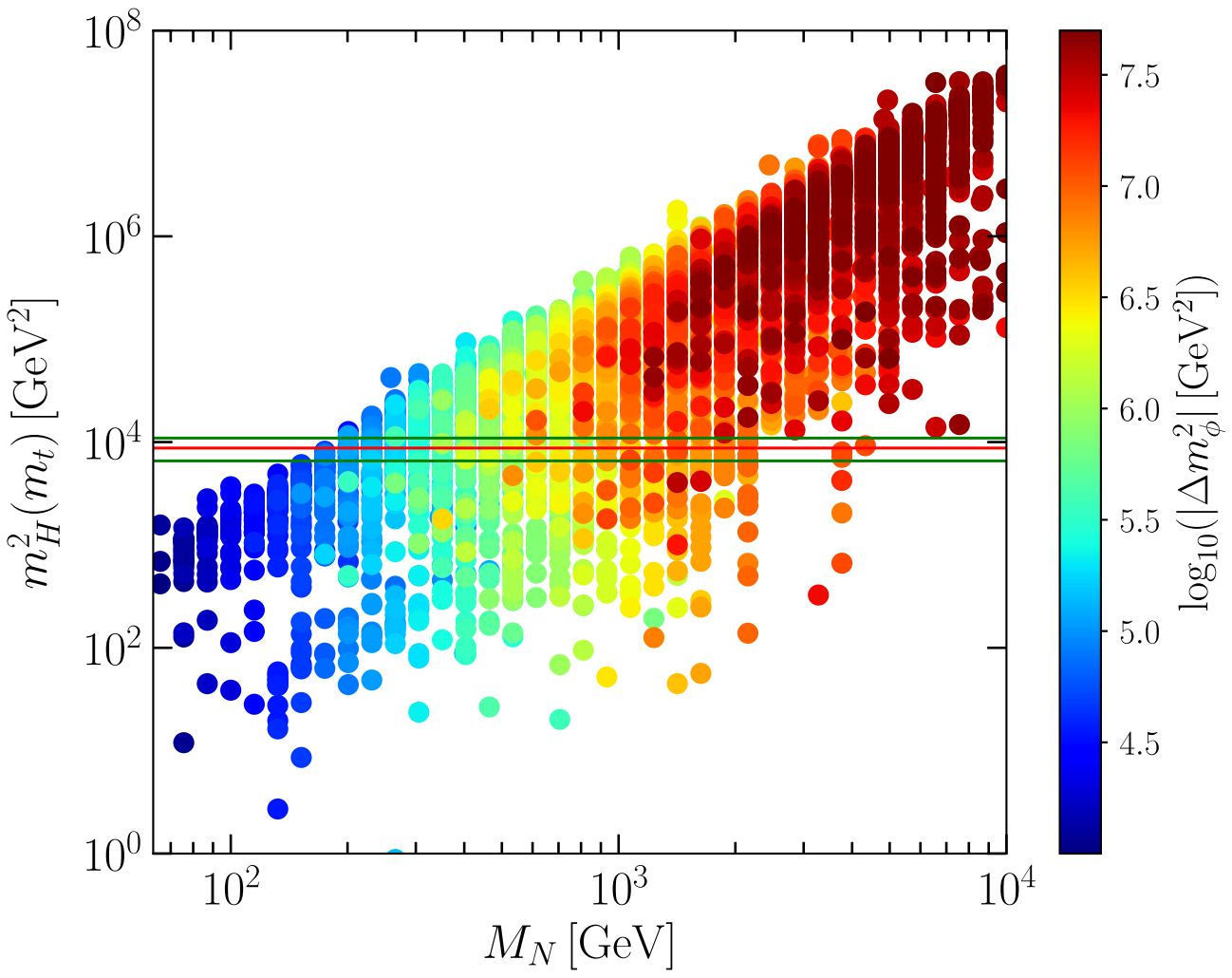

FIG. 1. This plot shows the radiatively induced Higgs mass as a function of RHN masses for different levels of $m_H^2$ fine-tuning. The green lines indicate the accepted range specified in Eq. (51).

Furthermore, we cannot rely on one-loop contributions to $\Delta\lambda$ because, as seen above [Eq. (49)], it results in the heaviest scalar having a mass at or below the TeV scale. Then, since the GW scale is usually of the same order as the heaviest particle in the model, having sizable one-loop $\Delta\lambda$ shifts in this case would require the GW scale to be very low lying, and the RG running of the Higgs mass from its initial negative value at the GW scale to its final positive value of $m_H^2 \sim \mathcal{O}(10^4)$ GeV$^2$ at the top mass scale would have to happen fast. Such fast running can only happen with large Yukawa couplings and small RHN masses, or large levels of fine-tuning. In other words, requiring sizable one-loop shifts in $\lambda$ and low fine-tuning also results in severe instabilities. Note that this result is quite general in the context of seesaw models [11,42].

In conclusion, stabilizing the Higgs potential in this scenario while inducing the correct Higgs mass parameter and Higgs self-coupling is not possible without large levels of fine-tuning. This motivates extending the model even further, in a way that relaxes the tension between the vacuum stability condition and low-lying GW scales or, equivalently, fast and unstable RG runnings. This possibility is discussed below in Sec. IV.

#### 5. Baryon asymmetry through leptogenesis

In this subsection, we explore the possibility of generating the baryon asymmetry of the Universe through leptogenesis in our model, which involves the inverse seesaw mechanism for generating SM neutrino masses with relatively low-scale RHNs. The inverse seesaw scenario has been the basis of many leptogenesis studies (see, e.g., [27,43,44]), which intend to explain the observed baryon asymmetry of the Universe [45],

$$\eta_B \approx 6 \times 10^{-10}. \tag{56}$$

The reason for this is mostly related to the nearly degenerate character of the pseudo-Dirac neutrinos, inherent to the inverse seesaw framework. This property of the RHNs opens the door to an alternative mechanism for generating the baryon asymmetry of the Universe, with RHN masses around the TeV scale.

More concretely, achieving the correct baryon asymmetry via *thermal leptogenesis* (for a review and original references, see, e.g., [46]) leads to a lower bound on the RHN masses [47],

$$M_N \gtrsim 10^9 \text{ GeV}. \tag{57}$$

A novel approach that allows one to circumvent this lower bound and to produce the correct baryon asymmetry with much lower RHN masses is provided by the so-called *resonant leptogenesis* [48,49]. In this framework, an enhancement of the generated $CP$ asymmetry is obtained when the small mass splitting between two RHNs is comparable to their decay width. Accordingly, the inverse seesaw framework is very suitable for generating a net baryon asymmetry in this way, since two pairs of nearly degenerate RHNs are obtained when $\mu/M_N \ll 1$.

Our UV model matches to an inverse seesaw framework, plus a light scalar, at energy scales below the Gildener-Weinberg scale; it is worthwhile checking whether the parameter space allows one to obtain the baryon asymmetry in Eq. (56). In other words, in this section we aim to find out whether the present model can produce the correct value of $\eta_B$ while satisfying all the scale-invariant conditions as well as perturbativity and correct IR observables, including the light neutrino masses (see Table III). Assuming that the lightest RHN pair is significantly lighter than the other one, most of the generated $CP$ asymmetry will come from the decays of the lightest pair. One can then focus on the out-of-equilibrium decay of this lightest pair $\{N_{R_1}, N_{R_2}\}$ depicted in Feynman diagrams shown in Fig. 2.

Note that, however, in the present model it is plausible that the singlet scalar $S$ is relatively light and present in the low-energy theory obtained by integrating out the heavy scalar $\phi$. In this case, other diagrams of the following form involving the light scalar $S$ will also contribute to the decays of the RHNs (see also [50,51]),

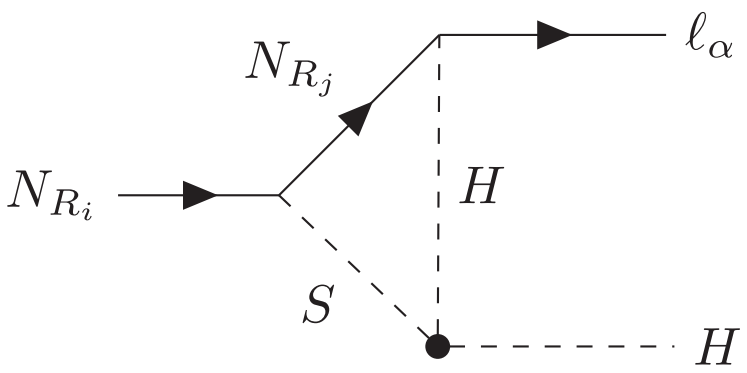

where the black dot represents an effective coupling, obtained after integrating out heavy scalar $\phi$. Furthermore, in Refs. [50,51], a type-I seesaw scenario with one extra scalar singlet was studied, and analogous diagrams with scalar mediators were shown, for part of the parameter space, to enhance the production of $CP$ asymmetry. This effect may result in successful leptogenesis for RHN mass scales even lower than $\mathcal{O}(1)$ TeV [50]. However, a numerical evaluation of the Boltzmann equations including the effects of the scalars as in [50,51] shows that for such enhancement to take place it is crucial that the trilinear coupling between the scalar singlet and the Higgs is large enough. This is shown in Fig. 3 where the parameter $\mu_*$ is defined as [51]

$$\mu_* \simeq \left(\frac{M_{N_1}^2}{M_{N_2} - M_{N_1}}\right)\left(\frac{5 \times 10^{-4}}{\alpha_{tr}}\right), \tag{58}$$

with $\alpha_{tr}$ the trilinear coupling between the RHNs and the scalar singlet. The quantity $\mu_*$ approximately gives the value of the trilinear $\mu$ coupling above which the washout

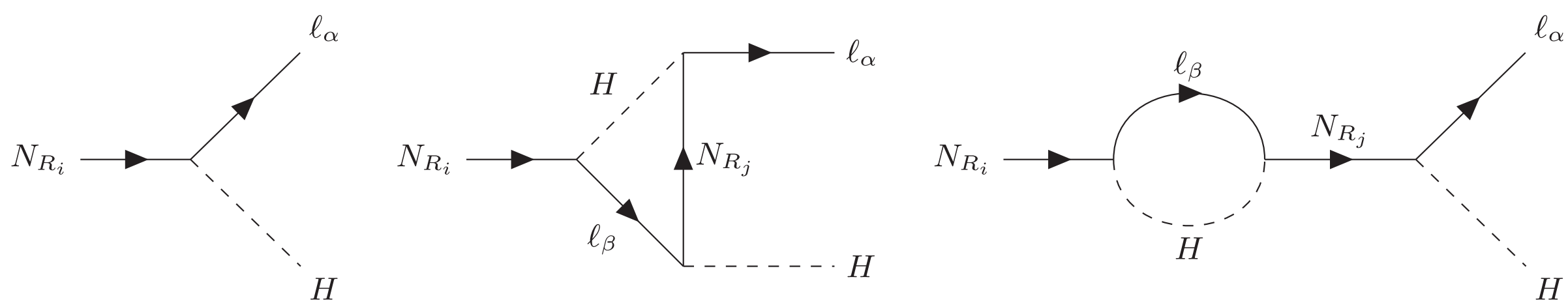

FIG. 2. Feynman diagrams contributing to the leptonic asymmetry in the inverse seesaw framework, where $i \neq j = 1, 2$ denote the two lightest RHN neutrinos.

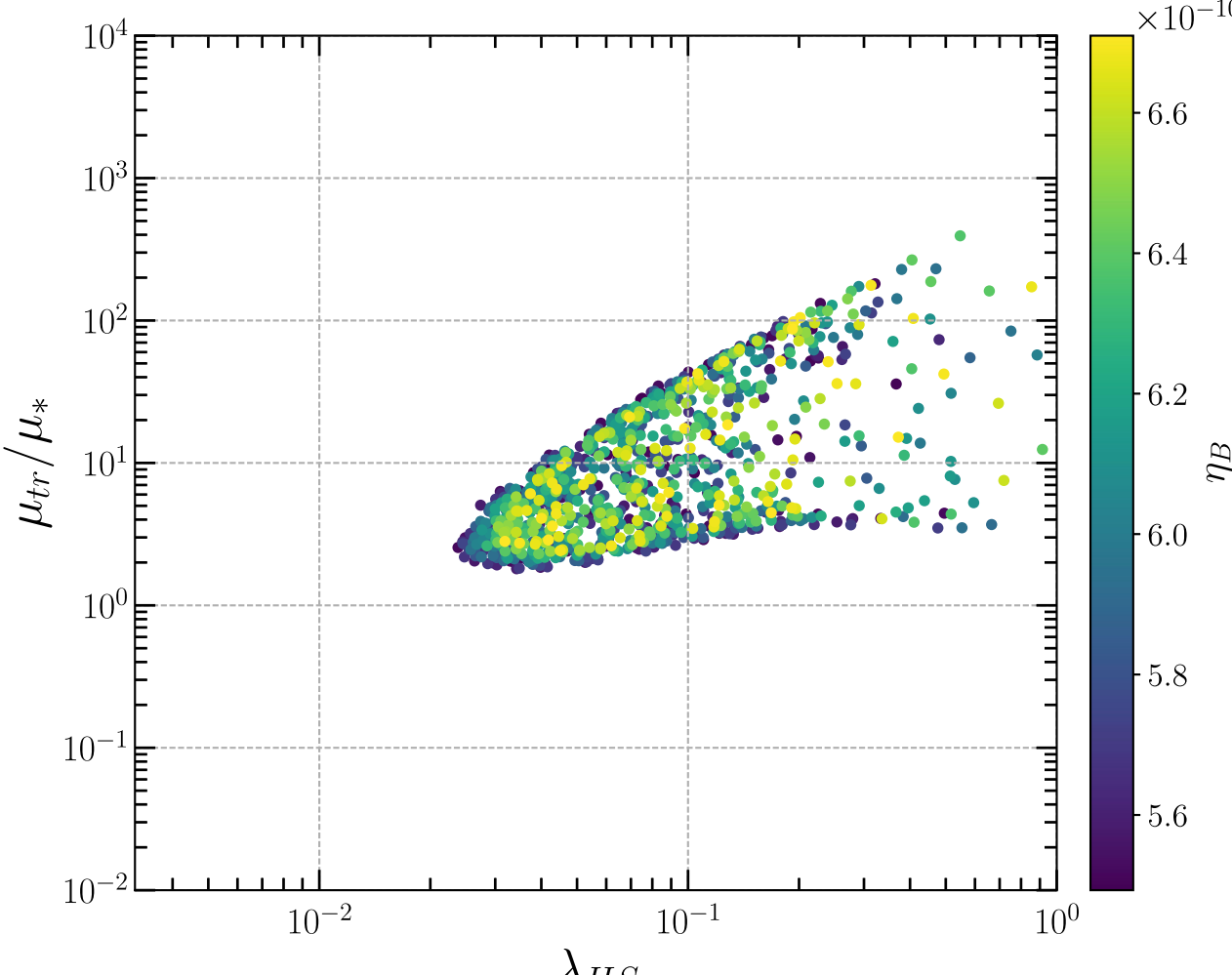

FIG. 3. Baryon asymmetry as a function of the Higgs-to-singlet trilinear coupling $\mu_{tr}$ and the Higgs-to-singlet portal coupling $\lambda_{HS}$, based on the numerical evaluation of the Boltzmann equations in [50,51].

effects due to the extra scalar singlet start to win over the enhancement effects.

In the current scenario, the trilinear coupling obtained after integrating out $\phi$ is given by

$$\mathcal{L}_{\rm eff} \supset \left(2\lambda_{HS} v_S + \frac{3}{2}\lambda_{H\phi} v_\phi \tan\alpha\right) S H^\dagger H, \tag{59}$$

implying that the value of the portal couplings $\lambda_{HS}$ and $\lambda_{H\phi}$ are relevant in determining the enhancement effects due to the singlet scalar. Figure 3 shows that the interesting values of the trilinear coupling require relatively large values of the portal coupling

$$\lambda_{HS} \gtrsim 10^{-2}, \tag{60}$$

which is inconsistent with the present scale-invariant scenario. Thus, we ignore this possible enhancement and focus on the pure inverse seesaw ingredients.

A full numerical evaluation involving RG solutions of the UV theory together with the solutions of the Boltzmann equations of the IR theory is out of the scope of the present study and will be the subject of future work. Instead, here we implement some of the approximate analytic results available in the literature for the computation of the baryon asymmetry in the inverse seesaw scenario [27,43].

The $CP$ asymmetry due to the decay of the first pseudo-Dirac pair is given by [27,43,44]

$$\begin{aligned}
\epsilon_1 &\approx \frac{1}{8\pi (h^\dagger h)_{11}} \mathrm{Im}[(h^\dagger h)^2_{12} f_{12} + (h^\dagger h)^2_{13} f_{13} + (h^\dagger h)^2_{14} f_{14}]\\
\epsilon_2 &\approx \frac{1}{8\pi (h^\dagger h)_{22}} \mathrm{Im}[(h^\dagger h)^2_{21} f_{21} + (h^\dagger h)^2_{23} f_{23} + (h^\dagger h)^2_{24} f_{24}],
\end{aligned} \tag{61}$$

where $f$, in the nearly degenerate case, is dominated by the self-energy contributions, i.e.,

$$f_{ij} \approx f^s_{ij} \approx \frac{M_j^2 - M_i^2}{(M_j^2 - M_i^2)^2 + |M_j \Gamma_j - M_i \Gamma_i|^2}. \tag{62}$$

In the above equation, $\Gamma_j$ represents the total decay rate of the $j$-th RHN and can be written as

$$\Gamma_j = \frac{M_j}{8\pi} (h h^\dagger)_{ii}. \tag{63}$$

Equation (62) and (61) make use of the $4\times 3$ matrix $h$ which for the case of diagonal $\mu$ and $M_N$ reads as [27]

$$\begin{aligned}
h_{1\alpha} &\simeq \frac{i e^{-i\theta_1/2}}{\sqrt{2}} \left(1 + \frac{\mu_1}{4 M_{N_1}}\right) y^*_{\alpha 1}\\
h_{2\alpha} &\simeq \frac{e^{-i\theta_1/2}}{\sqrt{2}} \left(1 - \frac{\mu_1}{4 M_{N_1}}\right) y^*_{\alpha 1}\\
h_{3\alpha} &\simeq \frac{i e^{-i\theta_2/2}}{\sqrt{2}} \left(1 + \frac{\mu_2}{4 M_{N_2}}\right) y^*_{\alpha 2}\\
h_{4\alpha} &\simeq \frac{e^{-i\theta_2/2}}{\sqrt{2}} \left(1 - \frac{\mu_2}{4 M_{N_2}}\right) y^*_{\alpha 2},
\end{aligned} \tag{64}$$

with $\alpha = 1, 2, 3$. In the weak and strong washout regimes, $\eta_B$ is approximately given by [43]

$$\eta_B = -\frac{28}{79} \begin{cases} \frac{\epsilon_1}{g_*} & K_{N_1^\pm} \ll 1,\\ \frac{0.3\epsilon_1}{g_* K_{N_1^\pm} (\ln K_{N_1^\pm})^{0.6}}, & K_{N_1^\pm} \gg 1, \end{cases} \tag{65}$$

with $g_* \approx 106.75$ the relativistic degrees of freedom of the SM, the washout parameter $K$, and the Hubble rate are defined as

$$K_{N_1^\pm} = \left.\frac{\Gamma_{N_1^\pm}}{2H(T)}\right|_{T=m_{N_1^\pm}} H(T) = \left(\frac{8\pi^3 g_*}{90}\right)^{\frac{1}{2}} \frac{T^2}{M_{\rm Pl}}. \tag{66}$$

Note that in the inverse seesaw, the Yukawa couplings are usually large, which results in very large washout parameters $K$. However, it has been shown that the relevant washout parameter $K_{\rm eff}$ is suppressed with respect to the $K$ defined above by a factor $\delta$ given by [27,52]

$$\delta \approx \frac{M_{N_2} - M_{N_1}}{\Gamma_1}. \tag{67}$$

Taking this into consideration and combining all of the above results, we plot in Fig. 4 the values for the baryon asymmetry obtained for a random sampling of points satisfying all scale-invariant, perturbativity, and low-energy constraints. This figure shows that the current scale-invariant framework may also be able to account for the observed baryon asymmetry of the Universe, favoring RHNs with mass in the $\mathcal{O}(10)$ TeV range. Furthermore, the expected resonant behavior is clear in this plot where a larger degeneracy in the mass of the pseudo-Dirac pair of RHNs, i.e., smaller $\delta$, leads to a larger baryonic asymmetry for a given mass $M_N$. For $\delta \sim 10^{-2}$ and within the reach of the above scan, the range of RHN masses leading to the correct order of $\eta_B$ is $M_N \sim 10$ TeV $-$ 50 TeV, while for $\delta \sim 10^{-3}$ this range shrinks and the overabundant solutions start to dominate. As already mentioned above, numerical computations involving the Boltzmann equations and the full system of RGEs are needed to make more reliable statements.

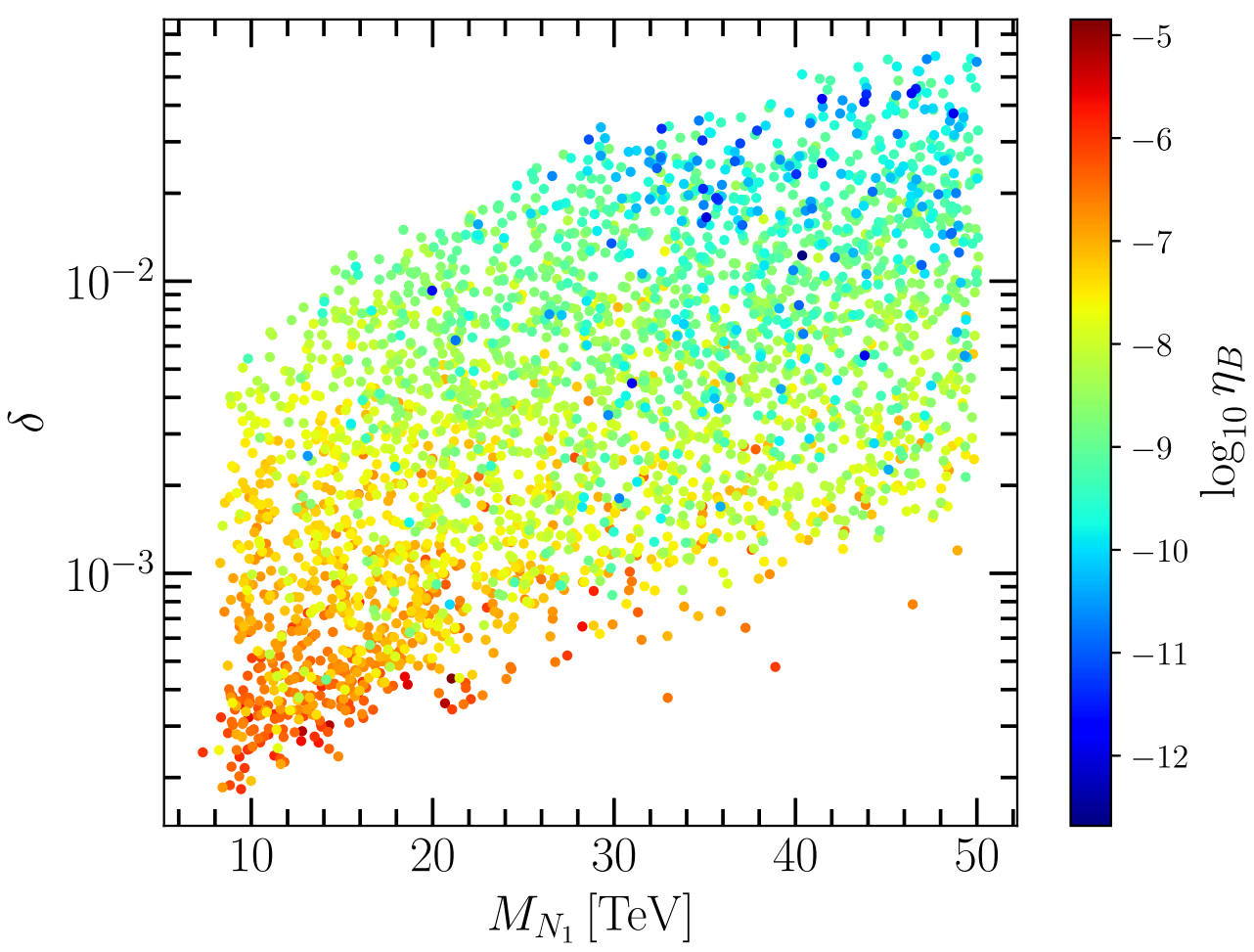

FIG. 4. Baryon asymmetry for UV parameters of the scale-invariant model satisfying all UV and IR constraints specified in Table III.

## IV. NEXT TO MINIMAL EXTENSION

In the scenario analyzed above, requiring that an appropriate minimum is generated for the one-loop effective potential and that the Higgs potential remains fully stable in the UV results in a great level of tension, and the model becomes inherently unstable/metastable. In this section, we provide a simple extension of the previous model to relax this tension without introducing additional fine-tuning.

The previous analysis shows that the heaviest scalar not only determines the scale of classical conformal symmetry breaking but is also responsible for the sizable shifts in the Higgs self-coupling. It was also shown that requiring larger $\Delta\lambda$ shifts pushes the heaviest scalar mass, and thus the GW scale, to lower values, resulting in severe instabilities. With this, our next main goal shall be to achieve sizable $\Delta\lambda$ shifts without having to rely on low-lying conformal symmetry-breaking scales, so that there is a large enough energy range for the $m_H^2$ top-down RG running to take place smoothly, i.e., without needing to rely on destabilizing large Yukawas. One way that this could be achieved is by having a heavy enough scalar in charge of setting the GW scale and a different one, which we will call $R$, responsible for generating a sizable shift in the Higgs self-coupling.

Thus, in this section, the original model presented above is extended by adding a third scalar, specifically a real scalar singlet, with a vanishing VEV at the GW scale, i.e., $\langle R\rangle = 0|_{\Lambda_{\mathrm{GW}}}$. One reason why the VEV of $R$ is chosen to vanish is that in this case the GW conditions of Eq. (7) and the field-dependent masses derived above remain unchanged. Another more important reason is that, with a vanishing VEV, $R$ will not contribute to the negative tree-level shifts $\Delta m^2_{\mathrm{tree}}$ of Eq. (20), and the portal $\lambda_{HR}$ can be sizable for fine-tuning $\Delta_{\mathrm{FT}} \gtrsim 0.1\%$. As mentioned above, a sizable portal is crucial to induce large enough shifts in the Higgs self-coupling. Finally, the introduction of this extra scalar with vanishing VEV will not introduce trilinear couplings with the Higgs bosons, and we therefore refrain from analyzing its impacts to the leptogenesis discussion above.

### A. Features of the extended scenario

The new scalar is assumed to have vanishing lepton number, and it will generally also couple to the RHNs, but since it does not acquire a nonvanishing VEV at $\Lambda_{\mathrm{GW}}$, it will not affect the inverse seesaw scenario obtained below the scale of classical conformal symmetry breaking. Therefore, one may focus on its contribution to the scale-invariant potential of Eq. (4), which gets modified to

$$V(H,S,\phi,R) = V(H,S,\phi) + \lambda_{HR}R^2(H^\dagger H) + \lambda_{SR}S^2R^2 + \lambda_{R\phi}R^2\phi^2 + \lambda_R R^4, \qquad (68)$$

where $V(H,S,\phi)$ is given in Eq. (4). Since the VEV of $R$ at $\Lambda_{\mathrm{GW}}$ vanishes, the flat direction from the original scenario or, equivalently, the GW conditions of Eq. (7), are unaffected. The tree-level field-dependent masses for $S$ and $\phi$ are also unaffected, and so are the corresponding threshold corrections. The field-dependent mass for $R$ is given by

$$M_R^2(h) = (M_R^0)^2 + \lambda_{HR}h^2, \qquad (69)$$

where $(M_R^0)^2 \equiv 2\lambda_{SR}v_S^2 + 2\lambda_{R\phi}v_\phi^2$. From Eqs. (B8) and (B9), the corresponding one-loop threshold corrections are given by

$$\Delta m_R^2 = -\frac{\lambda_{HR}}{16\pi^2}(2\lambda_{SR}v_S^2 + 2\lambda_{R\phi}v_\phi^2) \times \left[1 + 2\log\left(\frac{2\lambda_{SR}v_S^2 + 2\lambda_{R\phi}v_\phi^2}{Q^2}\right)\right], \qquad (70)$$

$$\Delta\lambda_R = \frac{1}{8\pi^2}\lambda_{HR}^2\left[\frac{3}{2} + \log\left(\frac{2\lambda_{SR}v_S^2 + 2\lambda_{R\phi}v_\phi^2}{Q^2}\right)\right]. \qquad (71)$$

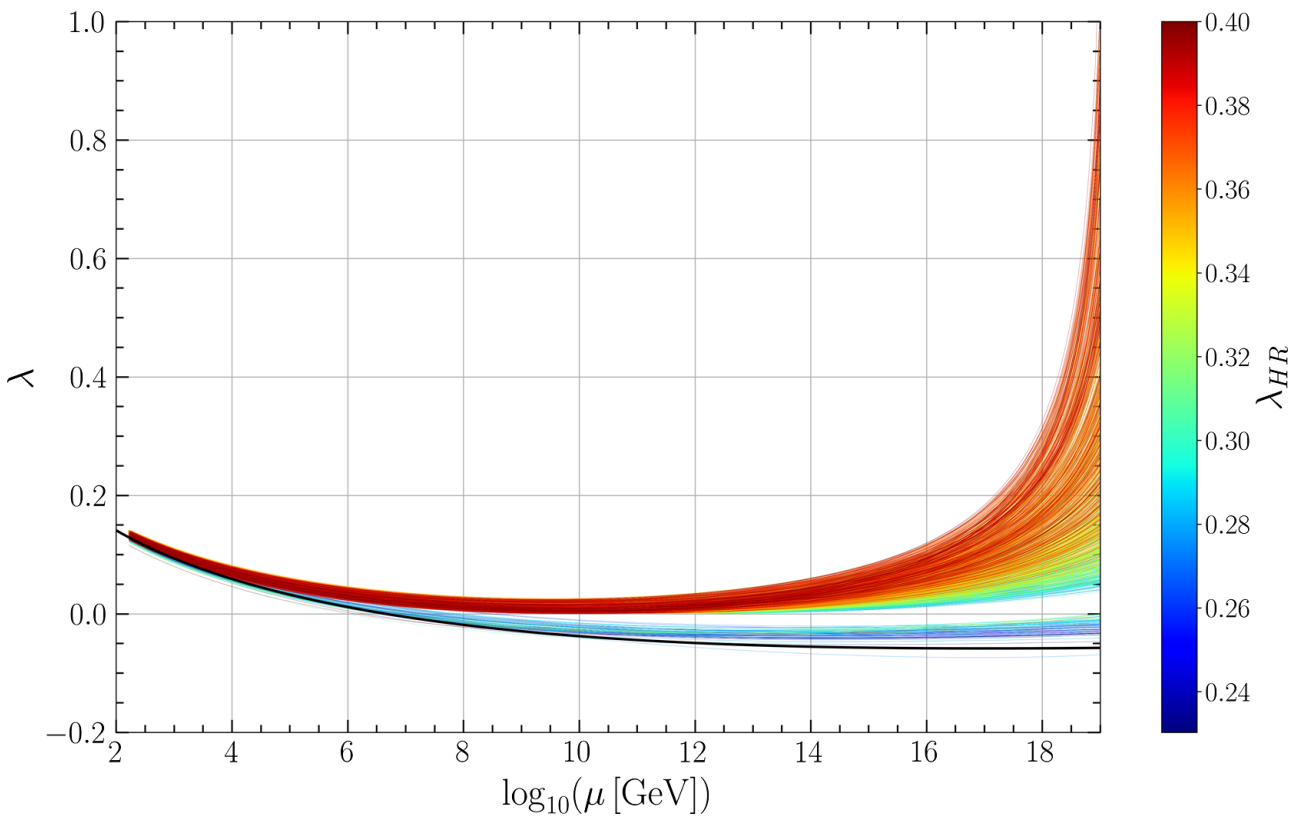

FIG. 5. RG running of the Higgs self-coupling in the extended scenario from the top mass scale to the Planck scale, as a function of the portal coupling $\lambda_{HR}$.

As seen above, increasing the mass of a given scalar effectively diminishes its contribution to the $\Delta\lambda$ shifts. In this scenario, the $B$ function is given by

$$B = \frac{1}{64\pi^2(v_S^2 + v_\phi^2)^2} \times \left(2|\Delta m^2_{\mathrm{tree}}|^2 + (M^0_\phi)^4 + (M^0_R)^4 - 8M^4_{N_2}\right). \tag{72}$$

Thus, by having $\phi$ massive enough to keep $B$ positive, $R$ can be light. The portal $\lambda_{HR}$ would, in principle, be bounded when suppressing its one-loop contribution to the mass-squared threshold correction as

$$|\Delta m^{2,R}_{\mathrm{loop}}| < \frac{m_H^2}{\Delta_{\mathrm{FT}}} \Rightarrow \lambda_{HR} < \frac{m_H^2}{(M^0_R)^2\Delta_{\mathrm{FT}}}. \tag{73}$$

However, the portal remains unsuppressed with no lower bound on $M_R$; the portal remains unsuppressed. Also, similarly to the result obtained in Eq. (49), when requiring sizable shifts in the Higgs self-coupling, the allowed masses for $R$, which now is the scalar responsible for allowing $\Delta\lambda > \mathcal{O}(10^{-3})$, are bounded from above. Meanwhile, the $\phi$ scalar's mass can still be large. We also find that as a consequence of $\phi$'s mass being able to remain large even after requiring sizable $\Delta\lambda$ shifts and without the need for large levels of fine-tuning as defined in Eq. (39), large GW scales are achievable in this scenario.

We note that the GW conditions are unchanged by the introduction of $R$; the set of GW relations that have to be checked in the UV so that no extra flat directions are developed is enlarged. The situation is now as described in the last part of Appendix B. More specifically, the set of equations that must be checked at each RG step when performing the bottom-up running from $\Lambda_{\mathrm{GW}}$ to $M_{\mathrm{Pl}}$ is given by (B14).

### B. Vacuum stability in the extended scenario

In the same spirit of the analysis carried out in the previous scenario, we perform a log-linear scan in this extended model and find that the same results are attainable regarding correct low-energy behavior, classical scale invariance, and perturbativity. In this section, we go one step further and demonstrate that achieving full stabilization of the Higgs potential in this new scenario is possible. This is achieved while inducing the correct Higgs mass and Higgs self-coupling at low energies, remaining classically conformal and perturbative up to the Planck scale, generating the correct light neutrino masses, and satisfying the experimental bounds associated with the branching ratios of LFV processes and heavy neutral lepton mixing. To this extent, the focus is driven to a sizable portal coupling between the $R$ and Higgs scalars, i.e., $\lambda_{HR} \sim \mathcal{O}(10^{-1})$.

Figure 5 shows the result of the RG running of the Higgs self-coupling for boundary conditions with a sizable $\lambda_{HR}$ portal that satisfies all of the above constraints. These solutions can be separated into two branches. The first one is characterized by a portal $\lambda_{HR}$ that is not large enough to keep the RG running of $\lambda$ positive definite between $m_t$ and $M_{\mathrm{Pl}}$. This turns out to be the case when $\lambda_{HR} \lesssim 0.3$. A second set of solutions is characterized by a portal in the range $0.3 \lesssim \lambda_{HR} \lesssim 0.4$, in which case the RG running of the Higgs self-coupling can remain positive and perturbative up to the Planck scale, rendering the Higgs scalar potential fully stable. Note that, generally, the larger the portal, the larger the value of $\lambda$ at the Planck scale. Thus, even if $\lambda_{HR} < 1$, if $\lambda_{HR}$ is too large, the model will run into Landau poles before reaching the Planck scale.

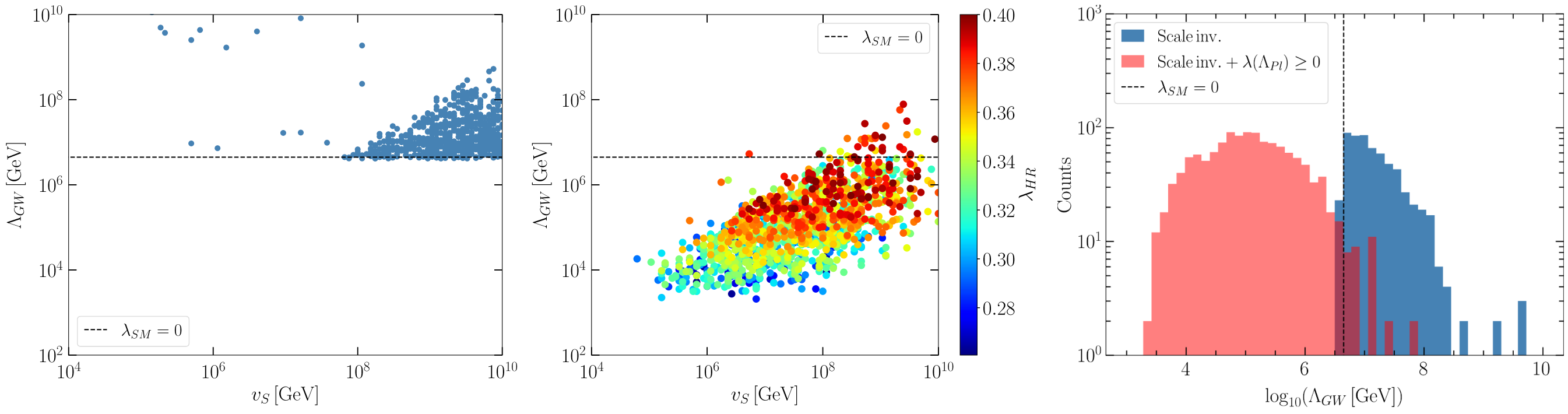

FIG. 6. GW scale as a function of the scalon VEV before (left) and after (middle) requiring full stabilization of the electroweak vacuum. The black dashed line indicates the energy scale at which the SM quartic coupling changes sign. The right panel shows the associated histograms.

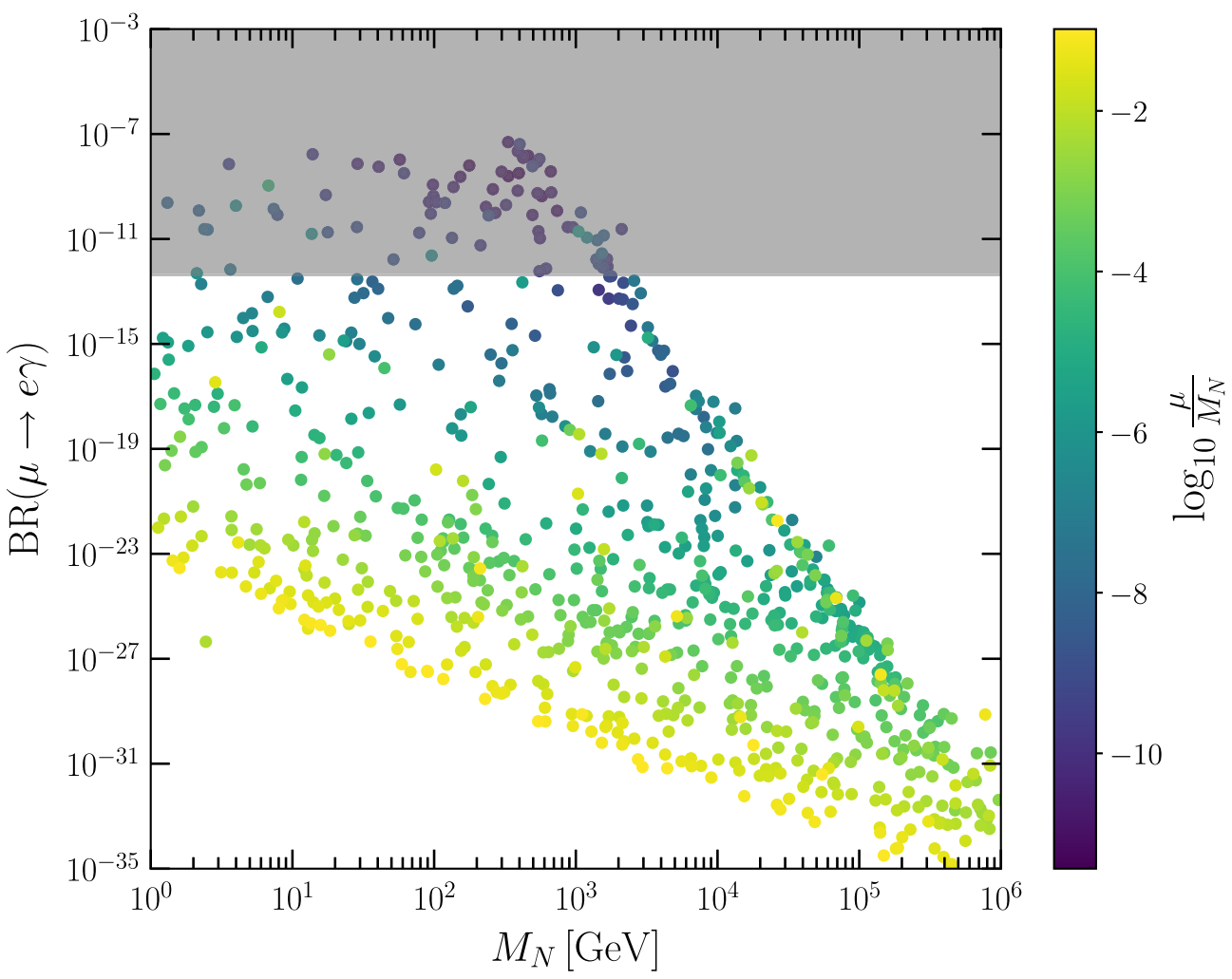

FIG. 7. RHN contributions to the branching ratio (BR) of the LFV process $\mu \to e + \gamma$ in the extended scenario as a function of the RHN mass. The shaded gray region corresponds to the experimental bound specified in Table II.

In Fig. 6, we show the GW scales associated with the two branches of solutions mentioned above. The first branch, shown in the left panel, consists of GW scales that are similar to or larger than the scale at which the SM Higgs self-coupling changes sign. The reason for this is that these solutions resemble that of the SM, i.e., they change sign at some scale below the Planck scale. Thus, their associated GW scales must be large enough so that this crossing happens at a high enough scale and no extra flat direction is developed between the GW scale and the Planck scale. The middle panel shows the GW scales associated with the second branch of solutions, where the Higgs scalar potential is rendered fully stable. Since in these cases the Higgs self-coupling does not change its sign in any step of the RG running, the classically conformal symmetry-breaking scale can be lower than the scale at which the SM Higgs self-coupling changes sign. There is also a clear dependency on the value of $\lambda_{HR}$, where larger values of the portal coupling allow for larger GW scales. This is because a larger $\lambda_{HR}$ induces a stronger "bending-up" effect in the bottom-up running of the Higgs self-coupling such that the classically conformal symmetry-breaking scale can happen at higher scales and still not cross the x axis. Finally, the right panel makes this behavior even more clear and shows that GW scales in the range $\mathcal{O}(10^4)$ GeV $< \Lambda_{\mathrm{GW}} < \mathcal{O}(10^6)$ GeV are favored when requiring the full stabilization of the Higgs scalar potential.

### C. Phenomenology

In this subsection, we briefly discuss the phenomenology of our model. First, we consider the branching ratios of LFV processes associated with points of the scan satisfying all low- and high-energy constraints in the extended scenario. For this, we run the involved couplings to the RHN scale, although this is not expected to change the results significantly due to their technically natural character. Figure 7 shows the result of a scan for the case $\mu \to e\gamma$. Since our model is flavor blind, the distributions for the $\tau \to \mu\gamma$ and $\tau \to e\gamma$ scenarios are very similar, only more suppressed due to the larger decay rate of the tau lepton. For RHN masses $M_N \lesssim 3$ TeV, some of the scan points are excluded by the current experimental bound. These correspond to points with small enough $\mu/M_N$ ratios. However, for $M_N \gtrsim 3$ TeV the branching ratios of LFV processes are suppressed enough to avoid the experimental constraint. Note that the ratio $\mu/M_N$ cannot be arbitrarily small if we want to reproduce the correct values for the light neutrinos while keeping the Yukawa couplings in the perturbative regime [see Eq. (18)]. However, a more stringent upper bound on this distribution comes from the allowed level of fine-tuning, as shown in the plot. Second, in Fig. 8, we show the scan points that satisfy all our model constraints discussed above in the presence of current experimental bounds on heavy neutral lepton mixings [12]. Even though not shown in the plot, the bounds for $M_{N_1} < 1$ GeV are very constraining, further justifying why this study focuses on the range $\mathcal{O}(1)$ GeV $< M_N < \mathcal{O}(10^6)$ GeV in the first place. Again, since our model is flavor blind, we only show the results for $\mu$ mixing, which is the most constrained one. Note that the mass range shown here includes those masses of RHNs potentially leading to the correct baryon asymmetry (see Fig. 4). The lower bound on this distribution comes from the constraint $\mu/M_N < 1$ required for the implementation of the inverse seesaw mechanism. As the $\mu/M_N$ ratio increases, the entries of the mixing matrix in Eq. (25) decrease, and so does the mixing between light and heavy neutrinos. Finally, Fig. 8 also shows that the inclusion of $R$ allows for a wider range of RHN masses within the allowed range of fine-tuning as compared to the minimal scenario of Fig. 1.

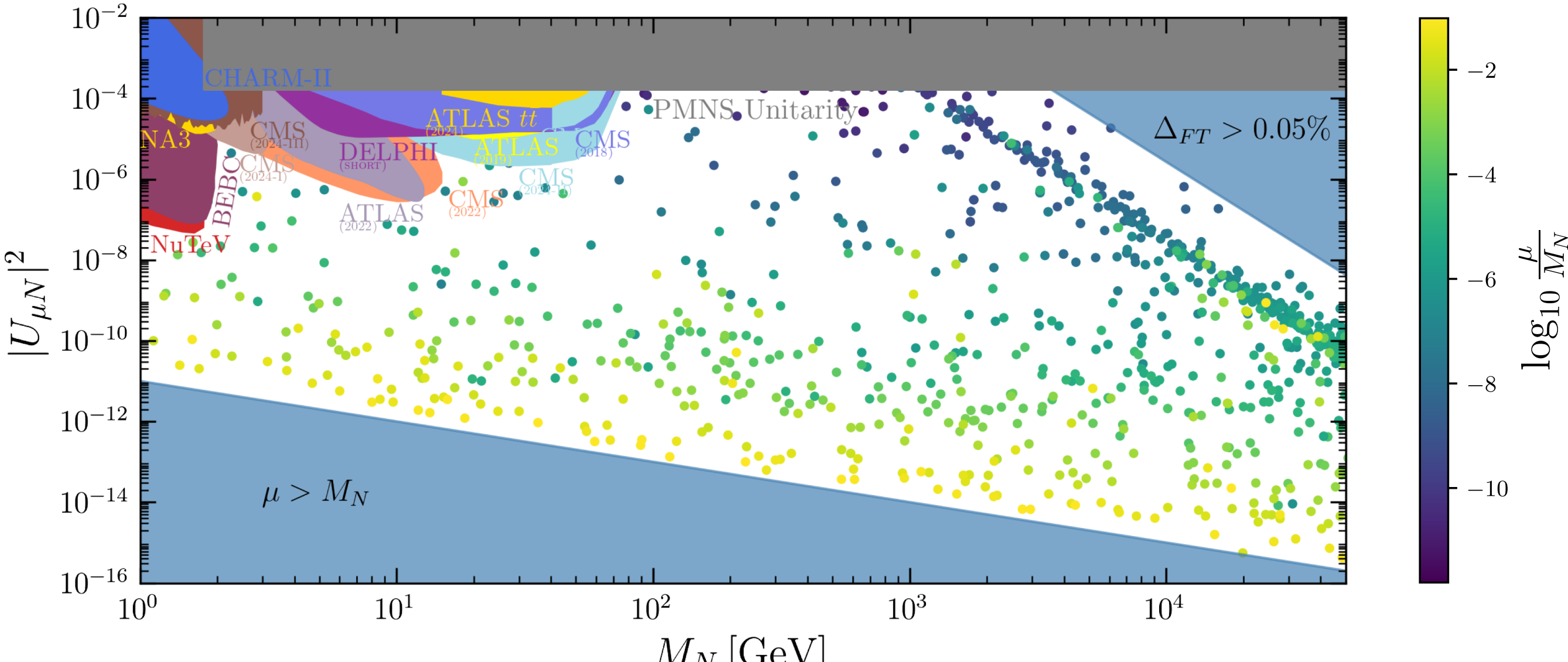

FIG. 8. Parameter points of a scan in the next-to-minimal extension scenario resulting in the correct Higgs mass, Higgs self-coupling, and light neutrino masses and satisfying the UV scale-invariant and perturbativity constraints up to $M_{\mathrm{Pl}}$. These are plotted together with the current experimental bounds on RHN mixings obtained from [12], where the theoretical bound due to unitarity of the Pontecorvo-Maki-Nakagawa-Sakata (PMNS) matrix applies above the tau lepton mass threshold. The color coding emphasizes the ratio of masses relevant to the mixing matrix in Eq. (25).

## V. SUMMARY AND CONCLUSIONS

While the SM can remain consistent up to very high energy scales, embedding it within a more fundamental framework typically introduces severe fine-tuning problems due to large quadratic corrections to the Higgs mass. Several approaches have been proposed to address this issue. One such proposal, known as the "neutrino option" [4], removes the explicit Higgs mass parameter from the Lagrangian and instead relies on RHNs to radiatively generate it. The absence of an explicit mass scale in the scalar sector motivated the development of a minimal conformal UV completion [7]. In that scenario, the SM is extended by two real scalar singlets and RHNs, with conformal symmetry broken à la Coleman-Weinberg, resulting in a type-I seesaw mechanism. However, to correctly reproduce the Higgs and light neutrino masses, RHN masses must lie in the 10–500 PeV range—far beyond current experimental reach.

This work explored alternatives to that minimal conformal realization. The first extension introduced one real and one complex singlet scalar, along with Weyl singlet fermions, realizing an inverse seesaw after classical scale symmetry breaking. We demonstrated that this model admits a classically conformal and perturbative UV completion up to the Planck scale, while also radiatively generating the correct Higgs mass. A key advantage is the significantly lower RHN mass scale. Specifically, we examined the range 1 GeV $< M_N < 10^6$ GeV and showed that conformal solutions exist across nearly this entire range. Importantly, for $M_N \lesssim 10^2$ GeV, current experimental bounds on heavy neutral lepton mixing apply. We found that parameter points with sufficiently large $\mu/M_N$ ratios remain within the allowed region. The same is true for bounds on LFV branching ratios, which are sensitive to the presence of RHNs.

After lowering the RHN mass scale to the testable regime, the next objective was to stabilize the Higgs potential. In the SM, the Higgs self-coupling becomes negative at high energies, rendering the potential metastable. This issue persists in the minimal conformal neutrino option, where the parameter space allows only highly suppressed shifts in the Higgs self-coupling. Although the first alternative in this study permits larger shifts, these remained insufficient for full stabilization. Specifically, achieving a sizable shift placed an upper bound on the new physics mass scales and implied a low GW scale. In this case, the RG evolution of the Higgs mass parameter from the initial shift (after integrating out the heaviest scalar) to the correct low-energy value needed to occur rapidly, requiring large Yukawa couplings. This, in turn, destabilized the RG running.

Setting aside the issue of vacuum stability, we investigated whether the model could generate the observed baryon asymmetry. We found that RHNs with masses of $\mathcal{O}(10)$ TeV are favored, though this depends nontrivially on the degeneracy parameter $\delta$. While a complete analysis was beyond the scope of this work, the model appears viable for resonant leptogenesis via the decay of pseudo-Dirac neutrinos and merits further study.

To address vacuum stability, we proposed a next-to-minimal extension in which a new scalar is responsible for the shift in the Higgs self-coupling, while a separate, heavier scalar sets the GW scale. Concretely, we added a real scalar singlet with a vanishing VEV at the symmetry-breaking scale. This extension preserved the overall structure of the scale-invariant framework but introduced sufficient flexibility to stabilize the potential. The threshold correction $\Delta\lambda_R$ to the Higgs quartic coupling is proportional to the portal coupling $\lambda_{HR}$. For stabilization, we required $\Delta\lambda_R \sim \mathcal{O}(10^{-2})$, which in turn implies $\lambda_{HR} \sim \mathcal{O}(10^{-1})$. We found that full stabilization was possible for $0.3 \lesssim \lambda_{HR} \lesssim 0.4$: smaller values gave an insufficient shift, while larger ones led to Landau poles. Stabilization also favored GW scales in the range $\mathcal{O}(10^4)$ GeV $< \Lambda_{\mathrm{GW}} < \mathcal{O}(10^6)$ GeV, with all viable points remaining consistent with LFV and mixing bounds.

The upper bound imposed on the heaviest scalar in the original framework (to ensure a sizable $\Delta\lambda$ shift) was transferred to the $R$ scalar in the improved scenario. Consequently, $m_R$ had to lie below 1 TeV for an initial shift $\Delta m_H^2 \sim -\mathcal{O}(10^4)$ GeV$^2$. The model allows flexibility in the choice of initial shift, GW scale, and Yukawa couplings, enabling RHN and scalar masses to span a broad range. Constraining these more tightly would require additional assumptions about these quantities.

In conclusion, this study successfully constructed a scale-invariant UV completion of the neutrino option that radiatively generates the correct Higgs mass and light neutrino masses, lowers the RHN mass scale into the testable regime, and enables full stabilization of the Higgs potential—all while maintaining perturbativity and classical conformal symmetry up to the Planck scale.

## ACKNOWLEDGMENTS

J. P. G. acknowledges funding from the International Max Planck Research School for Precision Tests of Fundamental Symmetries (IMPRS-PTFS).

## DATA AVAILABILITY

No data were created or analyzed in this study.

## APPENDIX A: LIGHT NEUTRINO MASSES AND MIXINGS

In this Appendix,[19] we provide additional information on the light SM neutrino masses and mixings. The Hermitian light neutrino mass block in Eq. (18) can be diagonalized with a unitary transformation $U \approx U_{\mathrm{PMNS}}$, where $U_{\mathrm{PMNS}}$ denotes the standard PMNS lepton mixing matrix [53,54]. In the two-right-handed-neutrino framework,[20] there is a single physical Majorana phase that cannot be rotated away. The central values of the mixing angles, $CP$ phases, and neutrino mass splittings for normal and inverted hierarchies (normal hierarchy (NH) and inverted hierarchy (IH), respectively) are summarized in Table V.

During our analysis, we employ a version of the Casas-Ibarra parametrization [58] adapted to the inverse seesaw framework [59]. Denoting the diagonal light neutrino mass matrix by $\hat{m}_\nu = U^T m_\nu U = \mathrm{diag}(m_1, m_2, m_3)$, and working in a basis where $\mu$ is diagonal, leads to

$$\begin{aligned} \hat{m}_\nu &= U^T M_D (M_N^{-1})^T X^T \hat{\mu} X M_N^{-1} M_D^T U \\ &= \underbrace{U^T M_D (\tilde{M}_N^{-1})^T \sqrt{\hat{\mu}}}_{\equiv K} \underbrace{\sqrt{\hat{\mu}} \tilde{M}_N^{-1} M_D^T U}_{K^T}, \end{aligned} \tag{A1}$$

where the following relations have been used:

$$\begin{aligned} &\hat{\mu} = X^* \mu X^\dagger = \mathrm{diag}(\mu_1, \mu_2), \\ &X^\dagger X = \mathbb{I}, \qquad \tilde{M}_N^{-1} \equiv X M_N^{-1}. \end{aligned} \tag{A2}$$

TABLE V. PMNS matrix parameters and neutrino mass splittings obtained from the NUFIT4.0 global fit [55,56] and Super-Kamiokande atmospheric neutrino data [57]. The quoted uncertainties correspond to the $3\sigma$ confidence range.

| | NH | IH |
|---|---|---|
| $\sin^2\theta_{12}$ | $0.310^{+0.040}_{-0.035}$ | $0.310^{+0.040}_{-0.035}$ |
| $\sin^2\theta_{23}$ | $0.582^{+0.042}_{-0.154}$ | $0.582^{+0.041}_{-0.149}$ |
| $\sin^2\theta_{13}$ | $0.02240^{+0.00197}_{-0.00196}$ | $0.02263^{+0.00198}_{-0.00196}$ |
| $\delta/[\mathrm{rad}]$ | $3.79^{+2.6}_{-1.43}$ | $4.89^{+1.24}_{-1.47}$ |
| $\Delta m_{21}^2/10^{-5}$ [eV$^2$] | $7.39^{+0.62}_{-0.6}$ | $7.39^{+0.62}_{-0.6}$ |
| $\Delta m_{31}^2/10^{-3}$ [eV$^2$] | $2.525^{+0.0097}_{-0.0094}$ | |
| $\Delta m_{32}^2/10^{-3}$ [eV$^2$] | | $2.512^{+0.0094}_{-0.0099}$ |

In the IS(2, 2) scenario, one of the light neutrino masses is necessarily zero. For a NH, this corresponds to $m_1 = 0$, while for an IH, $m_3 = 0$. The remaining two mass eigenvalues are then fully determined by the mass-squared splittings listed in Table V.

The matrix $K$ defined in Eq. (A1) is a $3 \times 2$ complex matrix subject to the following ten independent constraints:

$$K_{\alpha i} K_{\beta i} = \delta_{\alpha\beta} m_\alpha, \tag{A3}$$

which reduce the number of free real parameters to two. We denote these by $\gamma = \rho e^{i\beta}$, with the allowed ranges given by

$$0 \leq \rho \leq 2\pi, \qquad 0 \leq \beta \leq 2\pi. \tag{A4}$$

The upper bound on $\rho$ ensures that the Yukawa couplings, which are proportional to $\cos\gamma$ and $\sin\gamma$, remain in the perturbative regime.

## APPENDIX B: ONE-LOOP EFFECTIVE POTENTIAL AND THE GILDENER-WEINBERG FORMALISM

In this Appendix, we discuss the GW formalism [20], both in general and in the context of our model presented in the main text.

### 1. Gildener-Weinberg formalism

The GW formalism provides an analytical approximation for the minimization of a potential with multiple scalar bosons. It assumes that a flat direction develops at the GW scale, denoted $\Lambda_{\rm GW}$. This flat direction leads to the GW conditions, and radiative corrections subsequently induce a minimum along this direction, where the involved scalars acquire nontrivial VEVs. The one-loop effective potential along a flat direction $\vec{\varphi}_{\rm flat} = \vec{n}\varphi$ is given by [20]

$$V_{\rm eff}^{(1)}(\vec{n}\varphi) = A\varphi^4 + B\varphi^4 \ln\frac{\varphi^2}{\Lambda_{\rm GW}^2}, \tag{B1}$$

where $\vec{n}$ is a unit vector, and $\varphi$ acts as a radial coordinate in field space. The functions $A$ and $B$ depend on the number of real degrees of freedom $d_i$, the spin $s_i$, and the tree-level field-dependent masses evaluated along the flat direction at the GW scale for all particles in the model. Specifically,

$$A = \frac{1}{64\pi^2\langle\varphi\rangle^4}\sum_i (-1)^{2s_i} d_i m_i^4(\vec{n}\langle\varphi\rangle)\left(\ln\frac{m_i^2(\vec{n}\langle\varphi\rangle)}{\langle\varphi\rangle^2} - c_i\right), \tag{B2}$$

$$B = \frac{1}{64\pi^2\langle\varphi\rangle^4}\sum_i (-1)^{2s_i} d_i m_i^4(\vec{n}\langle\varphi\rangle), \tag{B3}$$

where the constants $c_i$ depend on the renormalization scheme. In the $\overline{\rm MS}$ scheme, $c_i = 5/6$ for gauge bosons and $c_i = 3/2$ for scalars and fermions. The extremum along the flat direction will correspond to a minimum if and only if $B > 0$. The value $\langle\varphi\rangle$ at which this extremum occurs is related to the GW scale as

$$\langle\varphi\rangle = \Lambda_{\rm GW}\exp\left(-\frac{1}{4} - \frac{A}{2B}\right). \tag{B4}$$

[19]Further details on the results cited in the following appendixes can be found in Ref. [41].

[20]This minimal scenario is denoted "IS(2,2)" since it involves two Weyl singlet fermions of each species.

Finally, the scalar boson associated with the flat direction $\vec{\varphi}_{\rm flat}$ will correspond to a pseudo-Goldstone boson with a vanishing tree-level mass and a one-loop mass given by

$$m_{\rm PGB}^2 = \left.\frac{d^2 V_{\rm eff}^{(1)}(\vec{n}\varphi)}{d\varphi^2}\right|_{\varphi=\langle\varphi\rangle} = 8B\langle\varphi\rangle^2. \tag{B5}$$

### 2. Field-dependent masses and threshold corrections

The one-loop threshold corrections to the Higgs parameters resulting from integrating out new physics particles can be computed from the corresponding shift in the one-loop effective potential [15]:

$$\begin{aligned}\Delta V &= \pm\frac{1}{32\pi^2}M^4(h)\ln\frac{M^2(h)}{Q^2}\\ &= \Delta V_0 - \frac{\Delta m^2}{2}h^2 + \frac{\Delta\lambda}{4}h^4 + \mathcal{O}(h^6),\end{aligned} \tag{B6}$$

where $M(h)$ represents the field-dependent masses of the particles being integrated out, and $Q$ is related to the scale $\bar{Q}$ in the $\overline{\rm MS}$ scheme via $\bar{Q} = e^{-3/4}Q$. The plus sign is used for bosons, and the minus sign for fermions. For a scalar field-dependent mass,

$$M^2(h) = M_0^2 + \kappa h^2 + M_2^{-2}h^4, \tag{B7}$$

the shift in the Higgs mass parameter and self-coupling up to one-loop level are given by

$$\Delta m^2 = \Delta m_{\rm tree}^2 - \frac{1}{16\pi^2}M_0^2\kappa\left(1 + 2\log\frac{M_0^2}{Q^2}\right), \tag{B8}$$

$$\begin{aligned}\Delta\lambda = \Delta\lambda_{\rm tree} + \frac{1}{8\pi^2}\bigg[&\kappa^2\left(\frac{3}{2} + \log\frac{M_0^2}{Q^2}\right)\\ &+ M_0^2 M_2^{-2}\left(1 + 2\log\frac{M_0^2}{Q^2}\right)\bigg].\end{aligned} \tag{B9}$$

### 3. Flat directions

A generic scale-invariant potential can be parametrized as

$$V = \frac{1}{24}f_{ijkl}\varphi_i\varphi_j\varphi_k\varphi_l, \tag{B10}$$

where $\{\varphi_i\}$ represents the coordinates in field space, and $f_{ijkl}$ are the coupling constants allowing for linear combinations of these coordinates. For this potential, a tree-level flat direction parametrized as $\vec{\varphi}_{\rm flat} = \vec{n}\varphi$, with $\vec{n}\cdot\vec{n} = 1$, develops at $\Lambda_{\rm GW}$ when

$$\left.\frac{dV}{d\vec{\varphi}}\right|_{\vec{\varphi}=\vec{n}\varphi,\Lambda=\Lambda_{\rm GW}} = 0. \tag{B11}$$

Additionally, since only differences in the potential are physically relevant, the flat direction may be chosen to lie at zero:

$$V(\vec{n}\varphi)|_{\Lambda_{\rm GW}} = 0. \tag{B12}$$

Equations (B11) and (B12) can be used to obtain an equation involving only the scalar couplings of the theory, which are known as the GW conditions. These conditions are typically written as

$$R(f)|_{\Lambda_{\rm GW}} = 0, \tag{B13}$$

and when satisfied, the scalar potential $V$ develops a flat direction.

Flat directions can be classified according to the number of fields with nonvanishing VEVs; for a more general discussion see Ref. [22]. We denote a flat direction with $N$ nonvanishing components as a type-$N$ flat direction. In the case of four scalars, relevant for the present analysis, there are four possible types of flat directions based on this classification. Substituting the flat directions into the potential and writing down the minimization conditions results in the GW relations. These relations must be checked for each RG step when performing bottom-up running from $\Lambda_{\rm GW}$ to $M_{\rm Pl}$, and are given by

$$
\begin{aligned}
&\lambda=\lambda_S=\lambda_\phi=\lambda_R=0,\\
&\lambda_{HS}^2-4\lambda\lambda_S=\lambda_{H\phi}^2-4\lambda\lambda_\phi=\lambda_{S\phi}^2-4\lambda_S\lambda_\phi=\lambda_{HR}^2-4\lambda\lambda_R=\lambda_{R\phi}^2-4\lambda_R\lambda_\phi=\lambda_{SR}^2-4\lambda_S\lambda_R=0,\\
&\lambda_{HS}^2\lambda_\phi+\lambda_{H\phi}^2\lambda_S+\lambda_{S\phi}^2\lambda-\lambda_{HS}\lambda_{H\phi}\lambda_{S\phi}-4\lambda\lambda_S\lambda_\phi=\lambda_{HR}^2\lambda_\phi+\lambda_{H\phi}^2\lambda_R+\lambda_{R\phi}^2\lambda-\lambda_{HR}\lambda_{H\phi}\lambda_{R\phi}-4\lambda\lambda_R\lambda_\phi=0,\\
&\lambda_{RS}^2\lambda_\phi+\lambda_{R\phi}^2\lambda_S+\lambda_{S\phi}^2\lambda_R-\lambda_{RS}\lambda_{R\phi}\lambda_{S\phi}-4\lambda_R\lambda_S\lambda_\phi=\lambda_{HS}^2\lambda_R+\lambda_{HR}^2\lambda_S+\lambda_{SR}^2\lambda-\lambda_{HS}\lambda_{HR}\lambda_{SR}-4\lambda\lambda_S\lambda_R=0,\\
&\lambda_{HS}^2\lambda_{R\phi}^2+\lambda_{H\phi}^2\lambda_{SR}^2+\lambda_{HR}^2\lambda_{S\phi}^2-4\lambda\lambda_S\lambda_{R\phi}^2-4\lambda\lambda_\phi\lambda_{SR}^2-4\lambda\lambda_R\lambda_{S\phi}^2-4\lambda_S\lambda_\phi\lambda_{HR}^2-4\lambda_S\lambda_R\lambda_{H\phi}^2\\
&\quad-4\lambda_\phi\lambda_R\lambda_{HS}^2+4\lambda\lambda_{S\phi}\lambda_{SR}\lambda_{R\phi}+4\lambda_S\lambda_{H\phi}\lambda_{HR}\lambda_{R\phi}+4\lambda_\phi\lambda_{HS}\lambda_{HR}\lambda_{SR}+4\lambda_R\lambda_{HS}\lambda_{H\phi}\lambda_{S\phi}\\
&\quad-2\lambda_{HS}\lambda_{S\phi}\lambda_{HR}\lambda_{R\phi}+16\lambda\lambda_S\lambda_\phi\lambda_R-2\lambda_{HS}\lambda_{H\phi}\lambda_{SR}\lambda_{R\phi}-2\lambda_{H\phi}\lambda_{HR}\lambda_{S\phi}\lambda_{SR}=0. \qquad \text{(B14)}
\end{aligned}
$$

## APPENDIX C: HEAVY NEUTRINO FIELD-DEPENDENT MASSES

To perform the above analysis, we block diagonalize the $7\times 7$ mass matrix:

$$
M=\begin{pmatrix} 0 & M_D & 0\\ M_D^T & 0 & M_N\\ 0 & M_N^T & \mu \end{pmatrix}. \qquad \text{(C1)}
$$

Without loss of generality, we assume $M_N$ to be real and diagonal, and for simplicity, we take $\mu$ to be a complex diagonal matrix. A first, nearly maximal, rotation of the lower-right $4\times 4$ block is then performed. Following the parametrization of [26], we define the rotation matrix $U_1$ as

$$
U_1=\begin{pmatrix} 1 & 0 & 0\\ 0 & \sqrt{1-\mathcal{B}\mathcal{B}^\dagger} & \mathcal{B}\\ 0 & -\mathcal{B}^\dagger & \sqrt{1-\mathcal{B}^\dagger\mathcal{B}} \end{pmatrix},
$$

$$
\mathcal{B}=\frac{1}{\sqrt{2}}-\left(\frac{b_1}{m_N}+\frac{b_2}{m_N^2}+\dots\right), \qquad \text{(C2)}
$$

where $m_N$ denotes the characteristic scale of $M_N$.

To handle the phases of $\mu$ *a priori*, we note that the $4\times 4$ bottom-right block of Eq. (C1) can be written as

$$
\begin{pmatrix} 0 & M_N\\ M_N^T & \mu \end{pmatrix}=\begin{pmatrix} D^* & 0\\ 0 & D^T \end{pmatrix}\begin{pmatrix} 0 & M_N\\ M_N^T & \mathrm{Re}(\mu) \end{pmatrix}\begin{pmatrix} D^\dagger & 0\\ 0 & D \end{pmatrix}, \qquad \text{(C3)}
$$

with

$$
\mu=\begin{pmatrix} \mu_{11}e^{i\theta_1} & 0\\ 0 & \mu_{22}e^{i\theta_2} \end{pmatrix}, \qquad D=\begin{pmatrix} e^{i\theta_1/2} & 0\\ 0 & e^{i\theta_2/2} \end{pmatrix}. \qquad \text{(C4)}
$$

The full rotation matrix is then defined as

$$
\tilde{U}_1=\begin{pmatrix} 1 & 0 & 0\\ 0 & D & 0\\ 0 & 0 & D^\dagger \end{pmatrix}\begin{pmatrix} 1 & 0 & 0\\ 0 & \sqrt{1-\mathcal{B}\mathcal{B}^\dagger} & \mathcal{B}\\ 0 & -\mathcal{B}^\dagger & \sqrt{1-\mathcal{B}^\dagger\mathcal{B}} \end{pmatrix}. \qquad \text{(C5)}
$$

From now on, we denote $\mu\equiv\mathrm{Re}(\mu)$. Expanding the rotated matrix up to $\mathcal{O}(1/m_N^2)$ and requiring the off diagonal blocks of the $2\times 2$ bottom-right submatrix to vanish yields the RHN masses up to $\mathcal{O}(m_N^{-1})$:

$$
M_N^-=-\left(M_N-\frac{\mu}{2}+\frac{M_N^{-1}\mu^2}{8}\right), \qquad \text{(C6)}
$$

$$
M_N^+=M_N+\frac{\mu}{2}+\frac{M_N^{-1}\mu^2}{8}. \qquad \text{(C7)}
$$

A second rotation is then applied to compute the field-dependent masses $M_N(h)$ up to $\mathcal{O}(h^2)$. The mass matrix takes the form

$$
\begin{pmatrix} 0_{3\times 3} & (\tilde{M}_D)_{3\times 4}\\ (\tilde{M}_D)^T_{4\times 3} & (\tilde{M}_N)_{4\times 4} \end{pmatrix}, \qquad \text{(C8)}
$$

which is block diagonalized similar to the type-I seesaw case. We again use the rotation matrix parametrization in Eq. (C2), now with $\mathcal{B}$ a $3\times 4$ matrix. Since $M_N\gg M_D$, this corresponds to a small rotation around $\theta=0$, and the eigenstates remain approximately unchanged. Thus, while we change the basis for the computation of field-dependent masses, we continue working in the original basis afterward.

In this setup, $\tilde{M}_D$ is a $3\times 4$ matrix and $M_N$ is a real diagonal $4\times 4$ matrix. We expand $\mathcal{B}$ as

$$
\mathcal{B}=\frac{b_1}{m_N}+\frac{b_2}{m_N^2}, \qquad \text{(C9)}
$$

and, following the same procedure as before with $U_2$ the rotation matrix centered at $\theta=0$, we obtain the field-dependent masses of the RHNs up to $\mathcal{O}(m_N^{-1})$:

$$
\tilde{M}'(h)=\begin{pmatrix} -M_N+\frac{\mu}{2}-\frac{\mu^2}{8M_N}-M(h) & 0\\ 0 & M_N+\frac{\mu}{2}+\frac{\mu^2}{8M_N}+M(h) \end{pmatrix}, \qquad \text{(C10)}
$$

with

$$
M(h)\equiv\frac{1}{2}M_N^{-1}M_D(h)^\dagger M_D(h)+\mathrm{H.c.} \qquad \text{(C11)}
$$